\documentclass[aps, prx, twocolumn, a4paper, reprint, floatfix, nofootinbib, longbibliography]{revtex4-1}

\usepackage[utf8]{inputenc}
\usepackage{epsfig}
\usepackage[T1]{fontenc}
\usepackage[english]{babel}
\usepackage[table]{xcolor}
\usepackage{t1enc}
\usepackage{graphicx}
\usepackage{amssymb}
\usepackage{amsmath}
\usepackage{relsize}
\usepackage{overpic}
\usepackage{bm}
\usepackage{hyperref}

\usepackage[normalem]{ulem}

\newcommand{\avg}[1]{{\left<#1\right>}}

\usepackage{mathtools}
\def\multiset#1#2{\ensuremath{\left(\kern-.3em\left(\genfrac{}{}{0pt}{}{#1}{#2}\right)\kern-.3em\right)}}

\usepackage{amsmath}

\usepackage{verbatim}
\usepackage{overpic}
\usepackage{booktabs}

\usepackage{pdfpages}

\begin{document}

\title{Hierarchical block structures and high-resolution model selection in large networks}

\author{Tiago P. Peixoto}
\email{tiago@itp.uni-bremen.de}
\affiliation{Institut f\"{u}r Theoretische Physik, Universit\"{a}t Bremen, Hochschulring 18, D-28359 Bremen, Germany}


\begin{abstract}
  Discovering and characterizing the large-scale topological features in
  empirical networks are crucial steps in understanding how complex
  systems function. However, most existing methods used to obtain the
  modular structure of networks suffer from serious problems, such as
  being oblivious to the statistical evidence supporting the discovered
  patterns, which results in the inability to separate actual structure
  from noise. In addition to this, one also observes a resolution limit
  on the size of communities, where smaller but well-defined clusters
  are not detectable when the network becomes large.  This phenomenon
  occurs not only for the very popular approach of modularity
  optimization, which lacks built-in statistical validation, but also
  for more principled methods based on statistical inference and model
  selection, which do incorporate statistical validation in a formally
  correct way.  Here we construct a nested generative model that,
  through a complete description of the entire network hierarchy at
  multiple scales, is capable of avoiding this limitation, and enables
  the detection of modular structure at levels far beyond those possible
  with current approaches. Even with this increased resolution, the
  method is based on the principle of parsimony, and is capable of
  separating signal from noise, and thus will not lead to the
  identification of spurious modules even on sparse
  networks. Furthermore, it fully generalizes other approaches in that
  it is not restricted to purely assortative mixing patterns, directed
  or undirected graphs, and \emph{ad hoc} hierarchical structures such
  as binary trees. Despite its general character, the approach is
  tractable, and can be combined with advanced techniques of community
  detection to yield an efficient algorithm that scales well for very
  large networks.
\end{abstract}

\maketitle

\section{Introduction}

The detection of communities and other large-scale structures in
networks has become perhaps one of the largest undertakings in network
science~\cite{newman_communities_2011, fortunato_community_2010}. It is
motivated by the desire to be able to characterize the most salient
features in large
biological~\cite{girvan_community_2002,ravasz_hierarchical_2002,fletcher_jr_network_2013},
technological~\cite{albert_internet:_1999,yook_modeling_2002} and social
systems~\cite{girvan_community_2002, zhao_community_2011,
  palla_uncovering_2005}, such that their building blocks become
evident, potentially giving valuable insight into the central aspects
governing their function and evolution. At its simplest level, the
problem seems straightforward: Modules are groups of nodes in the
network that have a similar connectivity pattern, often assumed to be
assortative, i.e., connected mostly among themselves and less so with
the rest of the network. However, when attempting to formalize this
notion, and develop methods to detect such structures, the combined
effort of many researchers in recent years has spawned a great variety
of competing approaches to the problem, with no clear, universally
accepted outcome~\cite{fortunato_community_2010}.

The method that has perhaps gathered the most widespread use is called
modularity optimization~\cite{newman_finding_2004} and consists in
maximizing a quality function that favors partitions of nodes for which
the fraction of internal edges inside each cluster is larger than
expected given a null model, taken to be a random graph. This method is
relatively easy to use and comprehend, works well in many accessible
examples, and is capable of being applied in very large systems via
efficient heuristics~\cite{clauset_finding_2004,
blondel_fast_2008}. However it also suffers from serious drawbacks. In
particular, despite measuring a deviation from a null model, it does not
take into account the statistical evidence associated with this
deviation, and as a result it is incapable of separating actual
structure from those arising simply of statistical fluctuations of the
null model, and it even finds high-scoring partitions in fully random
graphs~\cite{guimera_modularity_2004}. This problem is not specific to
modularity and is a characteristic shared by the vast majority of
methods proposed for solving the same
task~\cite{fortunato_community_2010}. In addition to the lack of
statistical validation, modularity maximization fails to detect clusters
with size below a given
threshold~\cite{fortunato_resolution_2007,lancichinetti_limits_2011},
which increases with the size of the system as $\sim\sqrt{E}$, where
$E$ is the number of edges in the entire network. This limitation is
independent of how salient these relatively smaller structures are, and
makes this potentially very important information completely
inaccessible. Furthermore, results obtained with this method tend to be
degenerate for large empirical networks~\cite{good_performance_2010},
for which many different partitions can be found with modularity values
very close to the global maximum. In these common situations, the method
fails in giving a faithful representation of the actual large-scale
structure present in the system.

More recently, increasing effort has been spent on a different approach
based on the statistical inference of generative models, which encodea
the modular structure of the network as model parameters~\cite{
  hastings_community_2006, garlaschelli_maximum_2008,
  newman_mixture_2007, reichardt_role_2007, hofman_bayesian_2008,
  bickel_nonparametric_2009, guimera_missing_2009,
  karrer_stochastic_2011,ball_efficient_2011, reichardt_interplay_2011,
  zhu_oriented_2014, baskerville_spatial_2011}.
This approach offers several advantages over a dominating fraction of
existing methods, since it is more firmly grounded on well-known
principles and methods of statistical analysis, which allows the
incorporation of the statistical evidence present in the data in a
formally correct manner. Under this general framework, one could hope to
overcome some of the limitations existing in more \emph{ad hoc}
methods, or at least make any intrinsic limitations easier to understand
in light of more robust
concepts~\cite{decelle_inference_2011,decelle_asymptotic_2011,
mossel_stochastic_2012, peixoto_parsimonious_2013}.  The generative
model most used for this purpose is the stochastic block
model~\cite{holland_stochastic_1983,fienberg_statistical_1985,
  faust_blockmodels:_1992, anderson_building_1992,
  hastings_community_2006, garlaschelli_maximum_2008,
  newman_mixture_2007, reichardt_role_2007, hofman_bayesian_2008,
  bickel_nonparametric_2009, guimera_missing_2009,
  karrer_stochastic_2011,ball_efficient_2011, reichardt_interplay_2011,
  zhu_oriented_2014, baskerville_spatial_2011}, which groups
nodes in blocks with arbitrary probabilities of connections between
them. This very simple definition already does away with the restriction
of considering only purely assortative communities, and accommodates
many different patterns, such as core-periphery structures and bipartite
blocks, as well as straightforward generalizations to directed
graphs. In this context, the detectability of well-defined clusters
amounts, in large part, to the issue of model selection based on
principled criteria such as minimum description length
(MDL)~\cite{rosvall_information-theoretic_2007,
peixoto_parsimonious_2013} or Bayesian model selection
(BMS)~\cite{daudin_mixture_2008, mariadassou_uncovering_2010,
moore_active_2011, latouche_variational_2012, come_model_2013}. These
approaches allow the selection of the most appropriate number of blocks
based on statistical evidence, and thus avoid the detection of spurious
communities. However, frustratingly, at least one of the limitations of
modularity maximization is also present when doing model selection,
namely, the resolution limit mentioned above. As was recently shown in
Ref.~\cite{peixoto_parsimonious_2013}, when using MDL, the maximum
number of detectable blocks scales with $\sqrt{N}$, where $N$ is the
number of nodes in the network, which is very similar to the modularity
optimization limit. However, in this context, this limitation arises out
of the lack of knowledge about the type of modular structure one is
about to infer, and the \emph{a priori} assumption that all
possibilities should occur with the same probability. Here we develop a
more refined method of model selection, which consists in a nested
hierarchy of stochastic block models, where an upper level of the
hierarchy serves as prior information to a lower level. This
dramatically changes the resolution of the model selection procedure,
and replaces the characteristic block size of $\sqrt{N}$ in the
nonhierarchical model by much a smaller value that scales only
logarithmically with $N$, enabling the detection of much smaller blocks
in very large networks. Furthermore, the approach provides a description
of the network in many scales, in a complete model encapsulating its
entire hierarchical structure at once. It generalizes previous methods
of hierarchical community detection~\cite{clauset_structural_2007,
clauset_hierarchical_2008, rosvall_multilevel_2011,
sales-pardo_extracting_2007,
ronhovde_multiresolution_2009,kovacs_community_2010,park_dynamic_2010},
in that it does not impose specific patterns such as dendograms or
binary trees, in addition to allowing arbitrary modular structures as
the usual stochastic block model, instead of purely assortative
ones. Furthermore, despite its increased resolution, the approach
attempts to find the simplest possible model that fits the data, and is
not subject to overfitting, and, hence, will not detect spurious modules
in random networks. Finally, the method is fully nonparametric, and can
be implemented efficiently, with a simple algorithm that scales well
for very large networks.

In Sec.~\ref{sec:model}, we start with the definition of the model and
then we discuss the model-selection procedure based on MDL. We then move
to the analysis of the resolution limit, and proceed to define an
efficient algorithm for the inference of the nested model, and we
finalize with the analysis of synthetic and empirical networks, where we
demonstrate the quality of the approach. We then conclude with an
overall discussion.

\section{Hierarchical Model}\label{sec:model}

The original stochastic block model
ensemble~\cite{holland_stochastic_1983,fienberg_statistical_1985,
faust_blockmodels:_1992, anderson_building_1992} is composed of $N$
nodes, divided into $B$ blocks, with $e_{rs}$ edges between nodes of
blocks $r$ and $s$ (or, for convenience of notation, twice that number
if $r=s$). Here, we differentiate between two very similar model
variants: 1. the edge counts $e_{rs}$ are themselves the parameters of
the model; 2. the parameters are the probabilities $p_{rs}$ that an edge
exists between two nodes of the respective blocks, such that the edge
counts $\avg{e_{rs}} = n_rn_sp_{rs}$ are constrained on average. Both
are equally valid generative models, and as long as the edge counts are
sufficiently large, they are fully equivalent (see
Ref.~\cite{peixoto_entropy_2012} and Appendix~\ref{app:bms}). Here, we
stick with the first variant, since it makes the following formulation
more convenient. We also consider a further variation called the
degree-corrected block model~\cite{karrer_stochastic_2011}, which is
defined exactly as the traditional model(s) above, but one additionally
specifies the degree sequence $\{k_i\}$ of the graph as an additional
set of parameters (again, these values can be the parameters themselves,
or they can be constrained on average~\cite{peixoto_entropy_2012}). The
degree-corrected version, although it is a relatively simple
modification, yields much more convincing results on many empirical
networks, since it is capable of incorporating degree variability inside
each block~\cite{karrer_stochastic_2011}. As will be seen below, it is,
in general, also capable of providing a more compact description of
arbitrary networks than the traditional version.

\begin{figure}[t]
  \includegraphics[width=0.95\columnwidth]{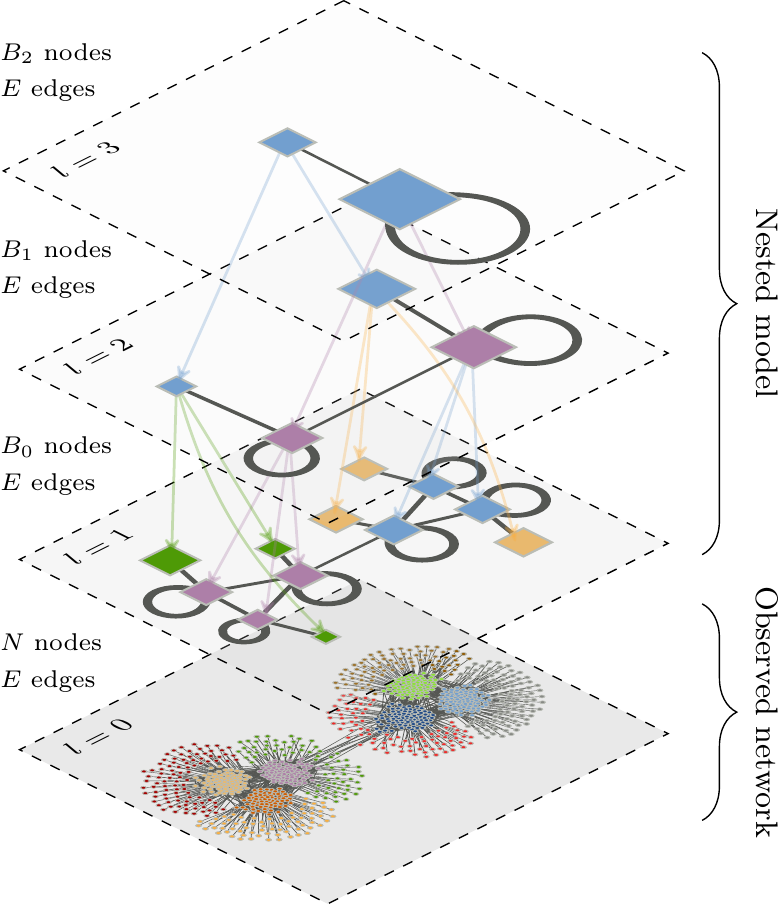}

  \caption{Example of a nested stochastic block model
  with three levels, and a generated network at the bottom. The
  top-level structure describes a core-periphery network, which
  is further subdivided in the lower levels.\label{fig:diagram}}
\end{figure}

The nested version, which we define, here is based on the simple fact that
the edge counts $e_{rs}$ themselves form a block multigraph, where the
nodes are the blocks, and the edge counts are the edge multiplicities
between each node pair (with self-loops allowed). This multigraph may
also be constructed via a generative model of its own. If we choose a
stochastic block model again as a generative model, we obtain another
smaller block multigraph as parameters at a higher level, and so on
recursively, until we finally reach a model with only one block. This
forms a nested stochastic block model hierarchy, which describes a given
network at several resolution levels (see
Fig.~\ref{fig:diagram}).

This approach provides an increased resolution when performing model
selection, since the generative model inferred at an upper level serves
as prior information to the one at a lower level.  Despite its more
elaborate formulation, this hierarchical model remains tractable, and it
is possible to apply it to very large networks, in a fully
nonparametric manner, as discussed below. Furthermore, it generalizes
cleanly the flat variants, which correspond simply to a hierarchy with
only one level. It also does not impose any preferred mixing pattern
(e.g., assortative or dissortative block structures), and is not
restricted to any specific hierarchical form, such as binary trees or
dendograms\footnote{This specification generalizes other hierarchical
constructions in a straightforward manner. For instance, the generative
model of Refs.~\cite{clauset_structural_2007,clauset_hierarchical_2008}
can be recovered as a special case by forcing a binary tree hierarchy,
terminating at the individual nodes, and a strictly assortative modular
structure. A similar argument holds for the variant of
Ref.~\cite{leskovec_kronecker_2010} as well.}. In the following, we
describe the maximum likelihood method to infer the multilevel
partitions, and the model selection process based on the minimum
description length principle, and compare it with Bayesian model
selection.

In the analysis, we focus on undirected networks, but everything is
straightforwardly applicable to directed networks as well. In
Appendix~\ref{app:summary} we present a summary of the relevant
expressions for the directed case.

\subsection{Module inference}

The inference approach consists in finding the best partition $\{b_i\}$
of the nodes, where $b_i \in [1, B]$ is the block membership of node
$i$, in the observed network $G$, such that the posterior likelihood
$\mathcal{P}(G|\{b_i\})$ is maximized. Since each graph with the same
edge counts $e_{rs}$ occurs with the same probability, the posterior
likelihood is simply $\mathcal{P}(G|\{b_i\}) =
1/\Omega(\{e_{rs}\},\{n_r\})$, where $e_{rs}$ and $n_r$ are the edge and
node counts associated with the block partition $\{b_i\}$, and
$\Omega(\{e_{rs}\},\{n_r\})$ is the number of different network
realizations. Hence, maximizing the likelihood is identical to
minimizing the ensemble entropy~\cite{peixoto_entropy_2012,
bianconi_entropy_2009} $\mathcal{S}(\{e_{rs}\},\{n_r\}) =
\ln\Omega(\{e_{rs}\},\{n_r\})$.

For the lowest level of the hierarchy (which models directly the
observed network), we have a simple graph, for which the entropies can be
computed as~\cite{peixoto_entropy_2012}
\begin{equation}\label{eq:st}
  \mathcal{S}_t = \frac{1}{2} \sum_{rs}n_rn_sH_{\text{b}}\left(\frac{e_{rs}}{n_rn_s}\right),
\end{equation}
for the traditional block model ensemble and,
\begin{equation}\label{eq:sc}
  \mathcal{S}_c \simeq -E -\sum_kN_k\ln k! - \frac{1}{2} \sum_{rs}e_{rs}\ln\left(\frac{e_{rs}}{e_re_s}\right),
\end{equation}
for the degree-corrected variant, where $E=\sum_{rs}e_{rs}/2$ is the
total number of edges, $N_k$ is the total number of nodes with degree
$k$, $e_r=\sum_se_{rs}$ is the number of half-edges incident on block
$r$, $H_{\text{b}}(x) = -x\ln x - (1-x)\ln (1-x)$ is the binary entropy
function, and it was assumed that $n_r \gg 1$. Note that only the last
term of Eq.~\ref{eq:sc} is, in fact, useful when finding the best block
partition, since the other terms remain constant. However the full
expression is necessary when comparing the models against each other via
model selection, as discussed below.

For the upper-level multigraphs the entropy can also be
computed~\cite{peixoto_entropy_2012}, and it takes a different form
\begin{equation}\label{eq:sm}
  \mathcal{S}_m  = \sum_{r>s} \ln{\textstyle \multiset{n_rn_s}{e_{rs}}} + \sum_r \ln{\textstyle \multiset{\multiset{n_r}{2}}{e_{rr}/2}},
\end{equation}
where $\multiset{n}{m}={n+m-1\choose m}$ is the number of
$m$-combinations with repetitions from a set of size $n$.  Note that we
no longer assume that $n_r\gg1$, since at the upper levels the number of
nodes becomes arbitrarily small.

At each level $l\in [0,L]$ in the hierarchy there are $B_{l-1}$ nodes,
which are divided into $B_{l}$ blocks (with $B_{l} \le B_{l-1}$), were
we set $B_{-1} \equiv N$.  The edge counts at level $l$ are denoted
$e^l_{rs}$, and the block sizes $n^l_r$. Therefore, we must have that
$\sum_rn^l_r=B_{l-1}$ and $\sum_{rs}e_{rs}^l/2 = E$; i.e., the total
number of nodes decreases in the upper levels, but the total number of
edges remains the same. The combined entropy of all layers is then given
by
\begin{equation}\label{eq:sn}
  \mathcal{S}_n  = \mathcal{S}_{t/c}(\{e^0_{rs}\}, \{n^0_r\}) + \sum_{l=1}^LS_m(\{e^l_{rs}\}, \{n^l_r\}).
\end{equation}
The full generative model corresponds to a nested sequence of network
ensembles, where each sample from a given level generates another
ensemble at a lower level. The entropy in Eq.~\ref{eq:sn} represents the
amount of information necessary to encode the decision sequence, which,
starting from the topmost model, selects the observed network among all
possible branches in the lower levels.

Whenever both the number of levels and the number of blocks $B_l$ of
each level is known, the best multilevel partition is the one that
minimizes $\mathcal{S}_n$. However, such information regarding the size
of the model is most often not available, and needs to be inferred from
the data as well. Using Eq.~\ref{eq:sn} for this purpose is not
appropriate, since minimizing it across all possible hierarchies leads
to a trivial and meaningless result where $B_l = N$ for all
$l$. Instead, one must employ some form of Occam's razor and select the
simplest possible model that best describes the observed data without
increasing its complexity. We present such an approach in the next
section.

\subsection{Model selection}

A method that directly formalizes Occam's razor principle is known as
minimum description length~\cite{grunwald_minimum_2007,
  rissanen_information_2010}, where one specifies the
\emph{total} amount of information necessary to described the data,
which includes not only the sample but the model parameters as well. The
description length for the model above is
\begin{equation}\label{eq:dl}
  \Sigma = \mathcal{L}_{t/c} + \mathcal{S}_{t/c},
\end{equation}
where $\mathcal{L}_{t/c}$ is the amount of information necessary to
fully describe the model, and $\mathcal{S}_{t/c}$ corresponds to entropy
of the lowest level $l=0$ of the hierarchy. In a given level $l$ of the
hierarchy, the information required to describe the model parameters
$\{e^l_{rs}\}$ is given by the entropy $S_m$ (Eq.~\ref{eq:sm}) of the
model in level $l+1$, so that we may write
\begin{equation}\label{eq}
  \mathcal{L}_{t} = \sum_{l=1}^LS_m(\{e^l_{rs}\}, \{n^l_r\}) + \mathcal{L}^{l-1}_t.
\end{equation}
The only missing information is how to partition the nodes of the
current level into $B_l$ blocks, which corresponds to the term
$\mathcal{L}^l_t$ in the equation above. The total number of partitions
with the same block sizes $\{n_r^l\}$ is given by
$B_{l-1}!/\prod_rn_r^l!$, and the total number of different block sizes
is $\multiset{B_l}{B_{l-1}}$. Hence, the amount of information necessary
to describe the block partition of level $l$ is
\begin{equation}\label{eq:dli}
  \mathcal{L}^l_t = \ln{\textstyle \multiset{B_l}{B_{l-1}}} + \ln B_{l-1}! - \sum_r \ln n_r^l!.
\end{equation}
Note that this is different from the choice made in
Refs.~\cite{peixoto_parsimonious_2013,
rosvall_information-theoretic_2007}, which considered all possible
$B_{l}^{B_{l-1}}$ partitions to be equally likely, and, hence, computed the
necessary amount of information as $B_{l-1}\ln B_l$. This choice
implicitly assumes that all blocks have approximately equal sizes, and
offers a worse description when this is not the case. Note that for
$B_{l-1} \gg 1$, we have
\begin{equation}
  \mathcal{L}^l_t \simeq B_{l-1} H(\{n_r^l / B_{l-1}\}),
\end{equation}
where $H(\{p_i\}) = -\sum_ip_i\ln p_i$ is the entropy of the
distribution $\{p_i\}$. Therefore, for uniform blocks $n^l_r = B_{l-1} /
B_l$ we recover asymptotically the value $\mathcal{L}^l_t \simeq
B_{l-1}\ln B_l$. However, the value of Eq.~\ref{eq:dli} can be much
smaller for nonuniform partitions. This choice has important
consequences for the resolution of relatively small blocks, as will be
seen below.

For the degree-corrected version, we still need to include the
information necessary to describe the degrees at the lowest level,
\begin{equation}\label{eq:l_c}
  \mathcal{L}_c =  \mathcal{L}_t + \sum_rn_rH(\{p^r_k\}),
\end{equation}
where $\{p^r_k\}$ is the degree distribution of nodes belonging to block
$r$~\footnote{Note that in Ref.~\cite{peixoto_parsimonious_2013} the
degree sequence entropy was taken to be $NH(\{p_k\})$, with
$p_k=\sum_rn_rp^r_k/N$, which implicitly assumed that the degrees are
uncorrelated with the block partitions, and hence should be interpreted
only as an upper bound to the actual description length given by
Eq.~\ref{eq:l_c}.}. It is worth noting that, if a network is sampled from
the traditional block model ensemble, so that $p_k^r$ is a Poisson with
average $e_r/n_r$, Eq.~\ref{eq:l_c} becomes $\mathcal{L}_c =
\mathcal{L}_t + 2E -\sum_re_r\ln e_r/n_r + \sum_kN_k\ln k!$, which means
that $\mathcal{S}_c + \mathcal{L}_c = \mathcal{S}_t + \mathcal{L}_t$,
i.e. the total description length is identical for both the traditional and
degree-corrected models in this case, and, therefore, both models describe
the same network equally well\footnote{Note that MDL can still be used
to select the simpler model in this case: Although the complete
description length $\Sigma$ will be asymptotically the same with both
models for networks sampled from the traditional block model, we still
have that $\mathcal{L}_t<\mathcal{L}_c$, since the degree-corrected
version still needs to include the information on the degree sequence,
as in Eq.~\ref{eq:l_c}.}. However, if the distributions $\{p^r_k\}$
deviate from Poissons, the degree-corrected variant will provide, in
general, a shorter description length.

It is easy to see that if one has a flat $L=1$ hierarchy, with $\{B_l\}
= \{B, 1\}$, the description length of the nonhierarchical model is
recovered~\cite{peixoto_parsimonious_2013}; e.g., for the traditional
model, we have $\Sigma_{L=1}= \mathcal{L}_{L=1} + \mathcal{S}_{t},$ with
\begin{equation}\label{eq:dl_flat}
  \mathcal{L}_{L=1}= \ln {\textstyle \multiset{\multiset{B}{2}}{E}} + \ln {\textstyle \multiset{B}{N}} + \ln N! - \sum_r \ln n_r!,
\end{equation}
where the only difference in comparison to
Ref.~\cite{peixoto_parsimonious_2013} is that here we are using the
improved partition description length of Eq.~\ref{eq:dli}. Therefore, the
nested generalization fully encapsulates the flat version, such that
$\min\Sigma \le \min\Sigma_{L=1}$; i.e., the nested model can provide
only a shorter or equal description length of the observed network.

The MDL principle predicates that whenever the hierarchy itself needs to
be inferred, one should minimize Eq.~\ref{eq:dl}, instead of
Eq.~\ref{eq:sn} directly. However, MDL is one of the many principled
methods one could use to do model selection, which include, e.g.,
Bayesian model selection via integrated
likelihood~\cite{biernacki_assessing_2000, hofman_bayesian_2008,
daudin_mixture_2008, mariadassou_uncovering_2010,
latouche_variational_2012, come_model_2013}, likelihood
ratios~\cite{yan_model_2012} or more approximative methods such as
Bayesian information criterion (BIC)~\cite{schwarz_estimating_1978} and
Akaike information criterion (AIC)~\cite{akaike_new_1974}. If any two of
these methods are derived from equivalent assumptions, one should expect
them to deliver compatible results. In Appendix~\ref{app:bms} we make a
comparison of the MDL approach with Bayesian model selection via
integrated likelihood (BMS), since it is nonapproximative and can be
computed exactly for the stochastic block model, where we show that
under compatible assumptions, these two methods deliver the exact same
results. In the following, we compare the results obtained with
nonhierarchical MDL/BMS and the nested model presented, and show that it
yields a higher quality model selection criterion, which detects the
correct number of blocks for sparse networks, without being
overconfident. Based on this analysis we are capable of deriving the
optimum number of blocks given a network size, and we show that the
nested model does not suffer from the resolution limit, which hinders the
nonhierarchical approach.

\subsubsection{Module detectability and the ``resolution limit''}

\begin{figure}
  \includegraphics[width=.9\columnwidth]{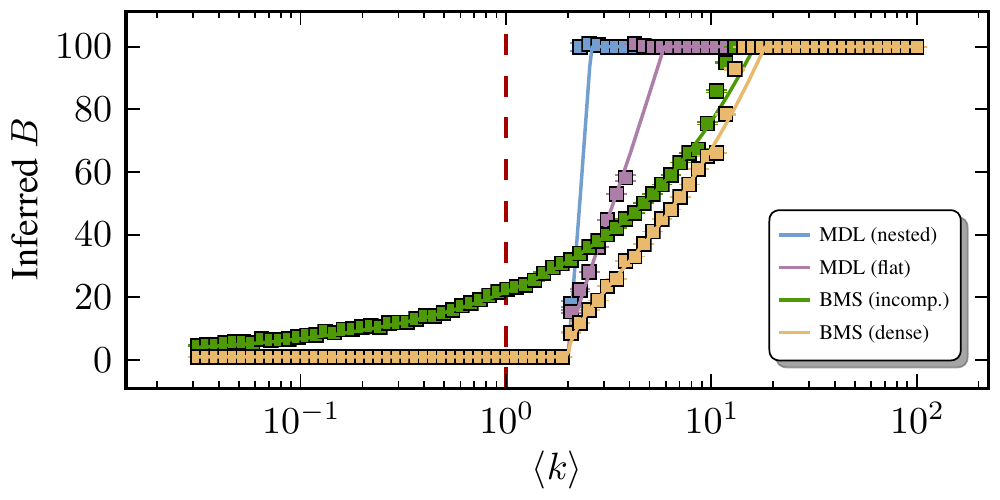}\\
  \includegraphics[width=.9\columnwidth]{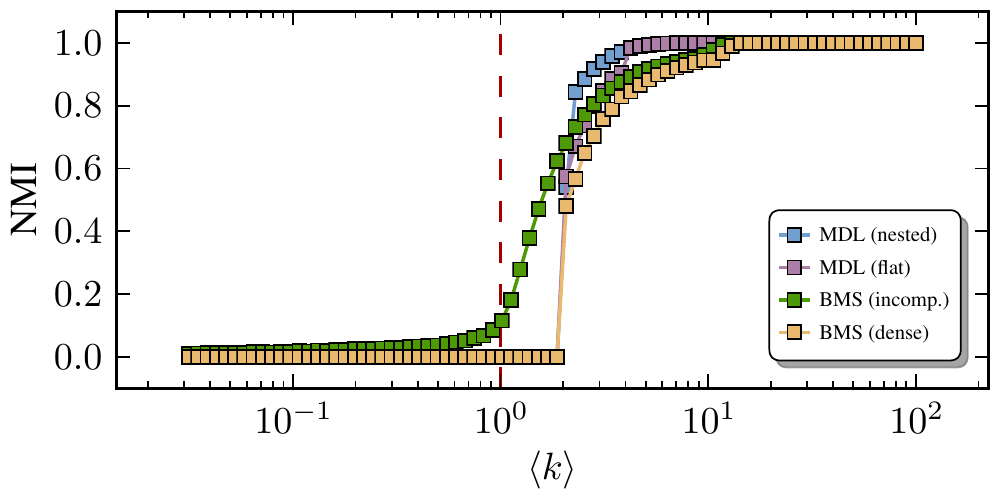}

  \caption{Model selection results for a PP model with $N=10^4$,
  $B_{\text{true}}=100$ and fully isolated blocks ($c=1$), using the
  model selection criteria described in the text. The top panel shows
  the inferred value of $B$ versus the average degree $\avg{k}$ in the
  network. The solid lines show the theoretical value according to each
  criterion, and the data points are direct optimization of the
  corresponding quantities for actual generated network, averaged over
  $40$ independent realizations. The bottom panel shows the normalized
  mutual information (NMI) between the inferred and planted partitions.
  The dashed line marks the threshold $\avg{k}=1$ where inference
  becomes impossible for $N\to\infty$. \label{fig:model_sel}}
\end{figure}

The general problem of module detectability can be formulated as
follows: Suppose we generate a network with a given parameter set. To
what extent can we recover the planted parameters by observing this
single sample from the model? The answer is conditional on the amount of
one's prior knowledge. If the number of blocks $B$ is known beforehand,
the remaining task is simply to classify the nodes in one of these $B$
classes. This problem has been shown to exhibit a
detectability-indetectability phase
transition~\cite{decelle_inference_2011, decelle_asymptotic_2011,
reichardt_detectable_2007, hu_phase_2012}: If the existing block
structure is too weak, it becomes impossible to infer the correct
partition with any method, despite the fact that the model parameters
deviate from that of a fully random graph. On the other hand, if the
block structure is sufficiently strong, it is possible to detect the
correct partition with a precision that increases as the block
structure becomes stronger. Another situation is when one does not know
the correct number $B$, which is arguably more relevant in practice. In
this case, in addition to the node classification, one needs to perform
model selection. Ideally, one would like to find the correct $B$ value
whenever the corresponding partition is detectable. However, in
situations where the correct partition is only \emph{partially}
detectable, i.e., the inferred partition is positively but weakly
correlated with the true model, an application of Occam's razor may
actually choose a simpler model, with smaller $B$, with a comparable
correlation with the true partition. Hence, if we lack knowledge of the
model size $B$, there will be situations where the true partition will
be more poorly detected, when compared to the case where we have this
information. This can be clearly illustrated with a very simple example
known as the planted partition (PP)
model~\cite{condon_algorithms_2001}. It corresponds to an assortative
block structure given by $e_{rs}=2E[\delta_{rs}c / B +
(1-\delta_{rs})(1-c) / B(B-1)]$, $n_r=N/B$, and $c\in[0, 1]$ is a free
parameter that controls the assortativity strength. For this model, if
we have that $N/B \gg 1$, it can be shown that the detectable phase
exists for $\avg{k} > [(B-1)/(cB-1)]^2$~\cite{decelle_inference_2011,
decelle_asymptotic_2011, mossel_stochastic_2012}. Let us make the
situation even simpler and consider the strongest possible block
structure with $c=1$, i.e. $B$ perfectly isolated assortative
communities with $N/B$ nodes. In this case the detectability threshold
lies at $\avg{k}=1$. Therefore, for any $\avg{k}>1$, we should be able
to detect all $B$ blocks, with a precision increasing with $\avg{k}$, if
we know we have $B$ blocks to begin with. If we do not know this, we must
apply a model-selection criterion as described above to obtain the best
value of $B$.  For simplicity, let us assume that, for the correct value
of $B\equiv B_{\text{true}}$ the true partition is perfectly detected,
such that $\mathcal{S}_t \simeq - E\ln B$, ignoring additive constants,
which are irrelevant at this point. If a value of $B>B_{\text{true}}$ is
used, we assume that the inferred partition corresponds to regular
subdivisions of the planted one, such that the entropy remains
approximately unchanged $\mathcal{S}_t \simeq - E\ln
B_{\text{true}}$. For $B<B_{\text{true}}$, the blocks are uniformly
merged together, so that $\mathcal{S}_t \simeq - E\ln B$. Hence, we may
write the expected value of the minimum description length in the
nonhierarchical model by summing $\mathcal{S}_t = -E
\ln\min(B,B_{\text{true}})$ with Eq.~\ref{eq:dl_flat}. For the nested
version of the model, we assume a regular hierarchy tree of depth $L$ and
with a fixed branching ratio $\sigma$, i.e. $B_l=\sigma^{L-l}$, so that
Eq.~\ref{eq:dl} becomes
\begin{multline}\label{eq:dl_opt}
  \Sigma \simeq \multiset{\sigma}{2} \frac{B}{\sigma-1}\ln E + \frac{\sigma}{2} B\ln B
  + N\ln B \\ - E\ln\min(B, B_{\text{true}}),
\end{multline}
where $B_l \gg \sigma$ was assumed, together with $L\gg 1$, and $B\equiv
B_0$. One may compare these criteria against each other in their
capacity of recovering the planted value of $B$, by finding the extremum
of each function. In Fig.~\ref{fig:model_sel}, we show the optimum
values of $B$ for a model with $N=10^4$ and $B_{\text{true}}=100$, as well as the
results for the direct minimization of the corresponding exact
quantities for actual network realizations, and a comparison of the
obtained partitions using the normalized mutual information (NMI)\footnote{The
normalized mutual information (NMI) is defined as
  $2I(\{b_i\},\{c_i\}) / [H(\{b_i\}) + H(\{c_i\})]$, where
  $I(\{b_i\},\{c_i\}) = \sum_{rs}p_{bc}(r,s) \ln \left(p_{bc}(r,
  s)/p_b(r)p_c(r)\right)$, and $H(\{x_i\}) = -\sum_rp_x(r)\ln p_x(r)$,
  where $\{b_i\}$ and $\{c_i\}$ are two partitions of the network.}. We
  also include the comparison with a dense BMS criterion
(see Appendix~\ref{app:bms}) both in its full form
(Eqs.~\ref{eq:il_summary} and \ref{eq:bms_dense}), and with the
partition likelihood term omitted, i.e. $\mathcal{P}(\{b_i\}|B) = 1$ in
Eq.~\ref{eq:il_summary}, as was done in
Refs.~\cite{guimera_missing_2009,moore_active_2011}. We see that the
dense BMS criterion fails to detect the correct model size for sparser
networks, which is in accordance with its inadequacy in this region. The
hierarchical model provides, as expected, the best results, and detects
the correct model for the sparsest networks. The incomplete BMS
criterion is clearly overconfident for sparse networks, and detects
$B>1$ structures even when the model lies below the detectability
threshold $\avg{k}=1$; hence, this shows that the partition likelihood
should not simply be discarded\footnote{The fact that the NMI between
the true and inferred partitions remains slightly above zero in
Fig.~\ref{fig:model_sel} for $\avg{k} < 1$ with the incomplete BMS
criterion is a finite size effect, as it tends increasingly to zero as
$N\to\infty$. On the other hand, according to this criterion, the
inferred value of $B$ in this region increases as $N$ becomes
larger. }. Both MDL and dense BMS fail to detect anything for $\avg{k} <
2$, which corresponds to a strong threshold\footnote{This threshold
corresponds simply the point where it becomes impossible to fully encode
the block partition in the network structure, i.e. for uniform blocks
$-E\ln B + N\ln B = 0$, which leads to $E=N$ and hence $\avg{k}=2$.},
which interestingly lies above the strict detectability limit at
$\avg{k}=1$. This corresponds to a region where detectability is
possible, but only if the true value of $B$ is known (or if a more
refined model-selection criterion exists). Note that the incomplete BMS
criterion performs better in the region $1<\avg{k}<2$, but this is
perhaps better interpreted as a by-product of its overall overconfidence
for very sparse networks. Note that all criteria eventually agree on the
correct value if $\avg{k}$ is made sufficiently large, which corresponds
to the intuitive notion that the detection problem becomes much easier
for dense networks.

A prominent problem in the detectability of block structures via other
methods, such as modularity optimization~\cite{newman_finding_2004} is
when modules are merged together, regardless of how strong the community
structure is perceived to be. For the modularity-based approach, when
considering a maximally modular network, similar to the PP model with
$c\to 1$, but with the additional restriction that the graph remains
connected, it has been shown~\cite{fortunato_resolution_2007} that
modules are merged together as long as $B > \sqrt{E}$. This phenomenon
is considered counterintuitive, and has been called the ``resolution
limit'' of community detection via this method\footnote{This limit
cannot be significantly changed even if one introduces scale parameters
to the definition of
modularity~\cite{lancichinetti_limits_2011,xiang_multi-resolution_2012}.}. As
it happens, this problem does not only occur for modularity-based
methods, but also if one does statistical inference based on MDL. For
the nonhierarchical model, it can be shown that, according to this
criterion, the optimal number of blocks scales as $B^* \simeq
\mu(\avg{k})\sqrt{N}$, where $\mu(x)$ is an increasing
function~\cite{peixoto_parsimonious_2013}. Therefore if the planted
number exceeds this threshold, blocks will be merged together, despite
the fact that the block structure is detectable with arbitrary precision
if one knows the correct value of $B$, and it sufficiently exceeds the
detectability threshold $\avg{k} > 1$ of the PP model. This means that
the true parameters of the model can no longer be used to compress the
generated data. This is a direct result of the assumption that all
possible block structures of a given size are equally possible, and the
number of such models becomes very large, with a model description
length scaling roughly with $\sim B^2\ln E + N\ln B$. In the presence of
additional assumptions about the model, such as the fact that one is
dealing with the PP model, instead of a more general block structure,
this can, in principle, be improved. However, in most practical situations
such assumptions cannot be made. One main advantage of the nested model
is that this limit can be overcome \emph{without} requiring such prior
knowledge. The description length via the nested model for the maximally
modular network above is given by Eq.~\ref{eq:dl_opt} with
$B_{\text{true}}=B$. As can be seen, this equation has only log-linear
dependencies on the model size $B$, instead of the quadratic one present
in the flat MDL. The result of this is that, if one finds the value of
$B^*$, which minimizes the nested description length, one obtains the
scaling
\begin{equation}
  B^* \propto  \frac{N}{\ln N},
\end{equation}
for sufficiently large $N$. This is a significant improvement, since the
maximum number of detectable blocks grows almost linearly with the
number of nodes. Thus, a characteristic detectable block size
$N/B^*\sim\sqrt{N}$ is replaced by a much smaller value $N/B^*\sim\ln N$,
which allows for a precise assessment of small communities even in very
large networks.

\begin{figure}
  \resizebox{.95\columnwidth}{!}{
    \hspace{-0.22cm}\includegraphics{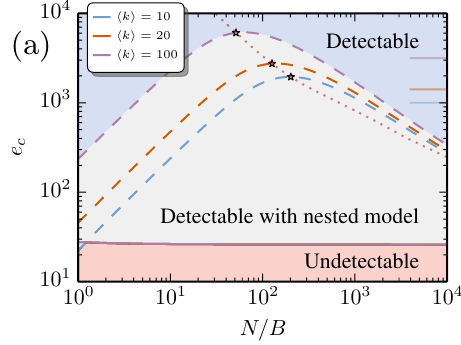}\includegraphics{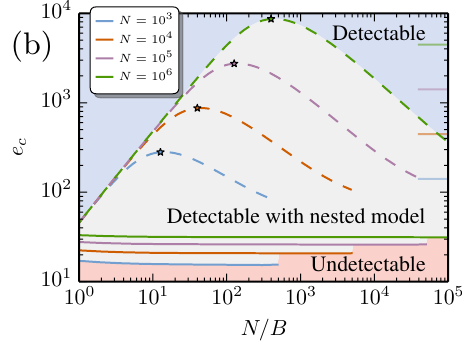}
  }
  \resizebox{.95\columnwidth}{!}{
    \includegraphics{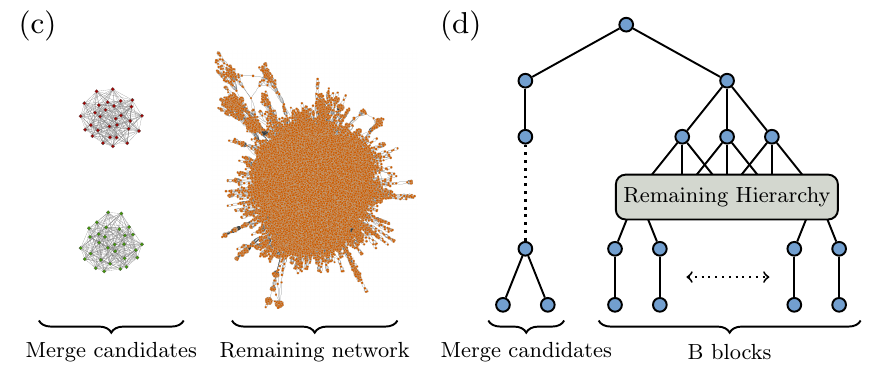}
  } \caption{Parameter region where two isolated blocks with $e_c/2$
  internal edges and $n_c=e_c/5$ internal nodes are detectable as
  separate blocks [shown schematically in panel (c)], as a function of
  the average block size $N/B$, and depending on (a) the average degree
  $\avg{k}$ with $N=10^5$ and (b) the number of nodes $N$ with
  $\avg{k}=20$.  The dashed curves show the boundaries for the
  nonhierarchical block model, and the solid lines for the hierarchical
  variant. The line segments on the right-hand side of the plots show
  the detectability threshold for
  modularity~\cite{fortunato_resolution_2007}, $e_c^* = \sqrt{2E}$.  The
  points marked with stars ($\star$) correspond to the maximum value of
  $B$ that is detectable in the remaining network with the
  nonhierarchical model, and the dotted line shows the same quantity
  for various $\avg{k}$ values (i.e. the region on the left of this
  curve corresponds to an overfitting of the remaining network,
  according to the nonhierarchical criterion).
  (d) The hierarchical construction used to decide if the
  two isolated blocks are merged together with the nested
  model. \label{fig:merging}}
\end{figure}

It is possible to understand in more detail the origin of the
improvement by considering a related problem, which is the detection of
specific blocks that are much smaller than the remaining
network. Another facet of the resolution limit manifests itself when two
such blocks are merged together, despite the fact that if they are
considered in isolation they would be kept separate. Here, we consider
this problem by using a slightly modified scenario than the one proposed
in Ref.~\cite{fortunato_resolution_2007}, which is a network composed of
two fully isolated blocks, each with $e_c/2$ internal edges and $n_c$
nodes, and a remaining network with $N$ nodes, $E$ edges, average degree
$\avg{k}=2E/N$ and an arbitrary topology [see
Fig.~\ref{fig:merging}(c)]. We may decide if these blocks are merged
together by considering the difference in the description length. The
entropy difference for the merge is simply $\Delta\mathcal{S}_t = e_c\ln
2$ (where we assume $e_c \ll n_c^2$, but the dense case can be computed
as well, with no significant difference in the result). For the flat
block model we have $\Delta\mathcal{L}_{\text{flat}} =
\mathcal{L}_{L=1}(E + e_c, N + 2n_c, B-1, \{n_r\} \cup \{2n_c\}) -
\mathcal{L}_{L=1}(E + e_c,N + 2n_c,B,\{n_r\} \cup \{n_c, n_c\})$,
computed using Eq.~\ref{eq:dl_flat}. For this case, the point at which
the merge happens, $\Delta\mathcal{L}_{\text{flat}} +
\Delta\mathcal{S}_t=0$, will depend not only on the values of $E$ and
$N$, but also on the average block size $N/B$ of the remaining network,
as can be seen in Fig.~\ref{fig:merging}(a) and (b). As the number of
blocks in the remaining network approaches the maximum detectable value,
$B^*\sim \sqrt{N}$, the more difficult it becomes to resolve the smaller
blocks. The detectable region recedes further with increasing $\avg{k}$,
and also with the number of nodes in the remaining network as $e_c^*
\sim \sqrt{N}$. Hence, the denser or larger the remaining network,
the harder it becomes to detect the smaller blocks with the flat variant
of the model. In Fig.~\ref{fig:merging} are also shown the values of
$e_r^*$ for which modularity also fails to separate the blocks (if one
considers that they are connected to themselves and to the rest of the
network by single edges\footnote{Note that in the model-selection
context, adding a single edge between the blocks is not a necessary
condition for the observation of the resolution limit, and has a
negligible effect, differently from the modularity approach, where it is
a deciding factor.}), which are overall compatible with the flat MDL
criterion. The situation changes significantly with the nested model. To
consider the merge, we assume an optimal block hierarchy which splits at
the top into two branches, the left one containing the two smaller
blocks, and the right one containing the remaining network and its
arbitrary hierarchical structure [see Fig.~\ref{fig:merging}(d)]. To
consider the merge, we need to compute the description length only at
the lowest level $l=0$, since the rest remains unchanged after the
merge. By computing the difference via Eq.~\ref{eq:dl}, after some
manipulations we obtain $\Delta\Sigma_{\text{nested}} =
\Delta\mathcal{S}_t + \ln n_c
- \ln\multiset{3}{e_c} + \ln(B+1) - \ln(B+N-1) - \ln(B_1 + B + 2)$,
with $B=B_0$. Note that this expression is independent of $E$, and, hence,
the density of the remaining network cannot influence the merging
decision. Since $B_1\leq B$, and assuming $B \gg 1$, we obtain
$\Delta\Sigma_{\text{nested}} \simeq \Delta\mathcal{S}_t + \ln n_c
- \ln[(e_c+2)(e_c+1)] -\ln(B + N)$, and, hence, the dependence on either
  $N$ or $B$ is again only logarithmic, $e^*_c \approx [\ln (B+N) - \ln n_c] / \ln 2$,
as shown in Fig.~\ref{fig:merging}(b). With this example one can notice
that the nested model is capable of compartmentalizing the network at
the upper levels, such that the lower-level branches can become almost
independent of each other. This means that, in many practical
situations, one can sufficiently overcome the resolution limit, without
abandoning a global model that describes the whole network at once.

In the following section we specify an efficient algorithm to infer the
parameters of the nested block model in arbitrary networks, and we test
its efficacy in uncovering the multilevel structure of synthetic as well
as empirical networks.

\section{Inference Algorithm}

Individually, any specific level $l$ of the hierarchical structure is a
regular block model, and, hence, the classification of the $B_{l-1}$
nodes of this level into $B_l$ blocks can be done via well-established
methods, such as the Monte Carlo method~\cite{moore_active_2011,
peixoto_parsimonious_2013}, simulated annealing, or belief
propagation~\cite{decelle_inference_2011, decelle_asymptotic_2011,
yan_model_2012}. Here, we use the method described in
Ref.~\cite{peixoto_efficient_2014}, which is an agglomerative heuristic
that provides high-quality results, while being unbiased with respect to
the types of block structure that are inferred, and is also very
efficient, with an algorithmic complexity of $O(N\ln^2 N)$, independent
of the number of blocks $B$. If one knows the depth $L$ of the
hierarchy, and all $\{B_l\}$ values, the multilevel partitions can be
obtained by starting from the lowest level $l=0$, and progressing
upwards to $l=L$. However, this cannot be done when the number and sizes
of the hierarchical levels are unknown. Although it is relatively simple
to heuristically impose such patterns as binary trees or dendograms,
these are not satisfactory given the general character of the model,
which accommodates arbitrary branching patterns. However, traversing all
possible hierarchies is not feasible for moderate or large networks;
thus, one must settle with approximative methods. Here, we propose a
very simple greedy heuristic, which, given any starting hierarchy,
performs a series of local moves to obtain the optimal
branching. Although this algorithm is not guaranteed to find the global
optimum, we have found it to perform very well for many synthetic and
empirical networks, and it tends to find consistent hierarchies,
independently of the starting estimate. It is also efficient enough to
allow its application to very large networks, since it does not
significantly change the overall algorithmic complexity of the inference
procedure. The algorithm is based on the following local moves at a
given hierarchy level $l$:
\begin{enumerate}
  \item {\bfseries Resize.} A new partition of the $B_{l-1}$ nodes into
        a newly chosen number of blocks $B_l$ is obtained. This is done
        via the agglomerative heuristic mentioned previously, with the
        modification that it must not invalidate the partition at the
        level $l+1$; i.e., no nodes that belong to different blocks at
        the upper level can be merged together in the current
        level. This restriction enables the difference in $\Sigma$
        (Eq.~\ref{eq:dl}) to be computed easily, since it depends only
        on the modifications made in the current and upper levels, $l$
        and $l+1$. The actual new value of $B_l$ is chosen via
        progressive bisection of the range $B_l \in [B_{l-1}, B_{l+1}]$,
        so that the minimum of $\Sigma$ is bracketed, and for each value
        of $B_l$ attempted, the best partition is found with the
        algorithm of Ref.~\cite{peixoto_efficient_2014}.
  \item {\bfseries Insert.} A new level is inserted at position $l$. Its
        size and partition are chosen exactly as in the resize move above.
  \item {\bfseries Delete.} The model in level $l$ is removed from the
        hierarchy; i.e., the nodes of level $l-1$ are grouped together
        directly as described in level $l+1$.
\end{enumerate}
Through repeated applications of these moves, it is possible to
construct any hierarchy. The actual greedy optimization consists of
starting with some initial hierarchy and keeping track of whether or not
each level is ``done'' or ``not done.'' One marks initially all levels
as not done and starts at the top level $l=L$. For the current level
$l$, if it is marked done it is skipped and one moves to the level
$l-1$. Otherwise, all three moves are attempted. If any of the moves
succeeds in decreasing the description length $\Sigma$, one marks the
levels $l-1$ and $l+1$ (if they exist) as not done, the level $l$ as
done, and one proceeds (if possible) to the upper level $l+1$, and
repeats the procedure. If no improvement is possible, the level $l$ is
marked as done and one proceeds to the lower level $l-1$. If the lowest
level $l=0$ is reached and cannot be improved, the algorithm ends. Note
that, in order to keep the description length complete, we must impose
that $B_L=1$ throughout the above process. The final hierarchy will, in
general, depend on the starting hierarchy, and as was mentioned above
one cannot guarantee that the global minimum is always found. However,
we find that, in the majority of cases, this algorithm succeeds in
finding the same or very similar hierarchies, independently of the
initial choice, which can simply be $\{B_l\} = \{1\}$. However, the
actual time it takes to reach the optimum will depend on how close the
initial tree was to the final one, and, hence, it is difficult to give
an estimate of the total number of moves necessary. However the slowest
move is the resize operation, which completes in $O(B_{l-1}\ln^2
B_{l-1})$ steps, and, hence, most of the time is spent at the lowest
level $l=0$ with $B_{-1}=N$, which scales well for very large
networks. We have succeeded in obtaining reliable results with this
algorithm for networks in excess of $10^7$ edges, hence it is suitable
for large-scale systems\footnote{An efficient and fully documented C++
implementation of the algorithm described here is freely available as
part of the graph-tool Python library at
\url{http://graph-tool.skewed.de}.}.

\section{Synthetic Benchmarks}

Here we consider the performance of the nested block model inference
procedure on artificially constructed networks. Here we use a nested
version of the usual PP model~\cite{condon_algorithms_2001}, inspired by
similar constructions done in
Refs~\cite{leskovec_kronecker_2010,palla_multifractal_2010}. We define a
seed structure with $B_0$ blocks and $[\bm{m}_1]_{rs}=\delta_{rs}c/B_0 +
(1-\delta_{rs})(1-c)/B_0(B_0-1)$, and construct a nested matrix of depth
$L-1$ via $\bm{m}_l = \bm{m}_{l-1} \otimes \bm{m}_{l-1}$ where $\otimes$
denotes the Kronecker product, and $l\in [1, L-1]$. The parameters of
the model at level $l$ are $e_{rs}^l= 2E m_{rs}$, and all $B=B_0^{L-1}$
blocks have the same number of nodes. Via spectral
methods~\cite{peixoto_eigenvalue_2013} one can show that the
detectability transition happens at $\avg{k} = [(B_0-1) / (cB_0-1)]^2$,
which is the same as the regular PP model with
$B=B_0$~\cite{decelle_inference_2011, decelle_asymptotic_2011,
mossel_stochastic_2012, nadakuditi_graph_2012}.

\begin{figure}
  \includegraphics[width=.9\columnwidth]{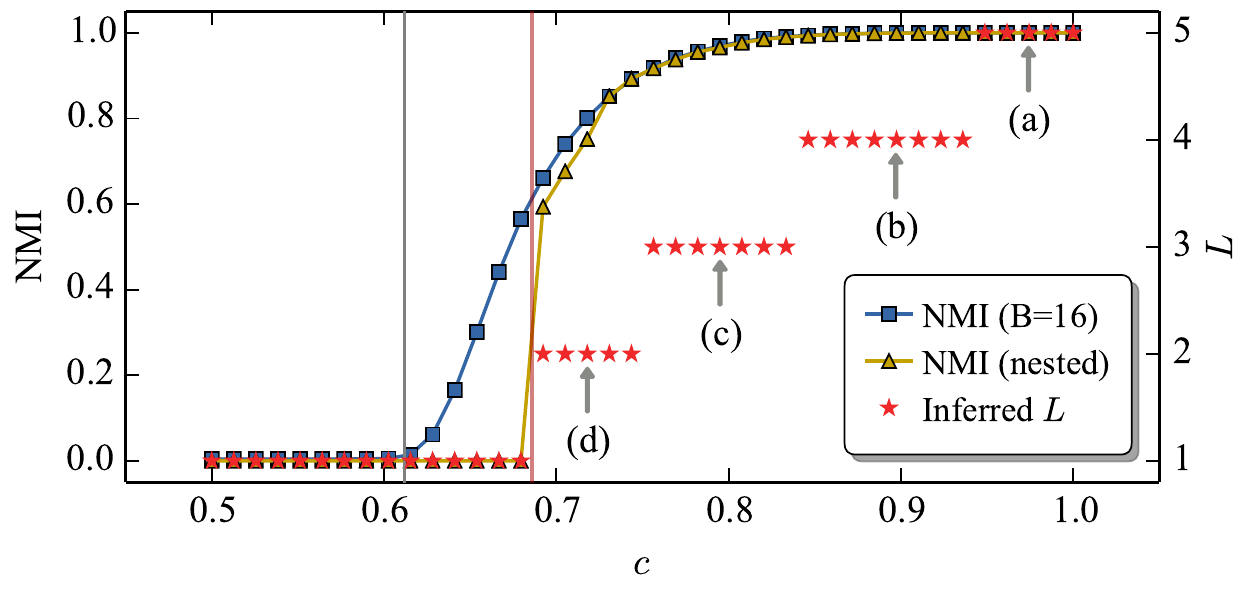}\\
  \begin{minipage}{.9\columnwidth}

    \includegraphics{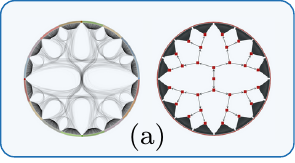}
    \includegraphics{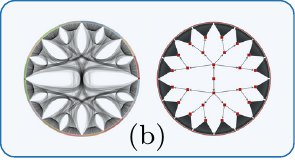}
    \includegraphics{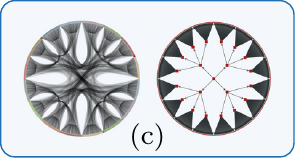}
    \includegraphics{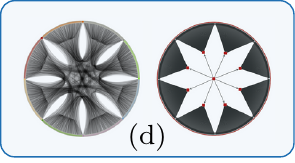}
  \end{minipage}

  \caption{\label{fig:nested-pp}Top: Normalized mutual information (NMI)
  between the inferred and true partitions for network realizations of
  the nested PP model described in the text with $B_1=2$, $L=5$,
  $\avg{k}=20$ and $N=10^4$, as a function of the assortativity strength
  $c$, both via the standard stochastic block model with $B=16$, and the
  nested variant with unspecified parameters. The star symbols ($\star$)
  show the value of $L$ for the inferred hierarchy. All points are
  averaged over $20$ independent realizations. The gray vertical line
  marks the detectability threshold $c^*$ when $B$ is predetermined, and
  the red line when the nested model fails to detect any
  structure. Bottom: Example hierarchies inferred for the values of $c$
  indicated in the top panel. The left image shows the network
  realization itself, and the right one the hierarchical structure [the
  planted hierarchy corresponds to the one in (a)].}
\end{figure}

\begin{figure*}
  \includegraphics[width=.48\textwidth]{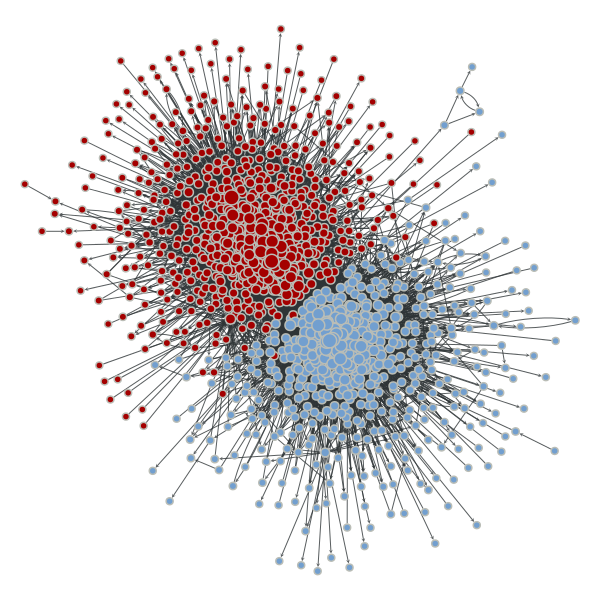}
  \includegraphics[width=.48\textwidth]{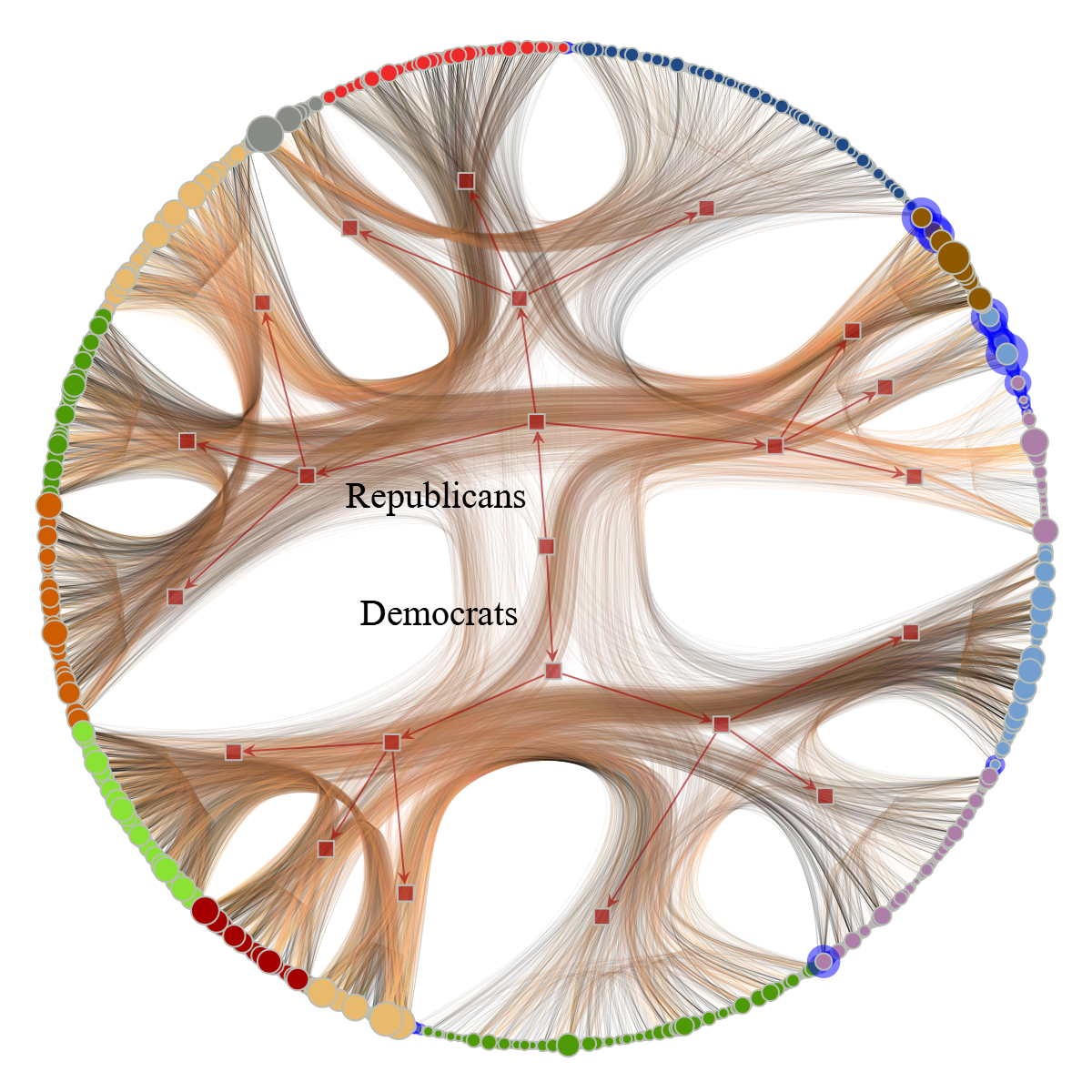}

  \caption{\label{fig:polblogs} The political blog network of Adamic and
  Glance~\cite{adamic_political_2005}. Left: Topmost partition of the
  hierarchy inferred with the nested model. Right: The same network,
  using a circular layout, with edge bundling following the inferred
  hierarchy~\cite{holten_hierarchical_2006} (indicated also by the
  square nodes, and the node colors). The size of the nodes corresponds
  to the total degree, and the edge color indicates its direction (from
  dark to light). Nodes marked with a blue halo were incorrectly
  classified at the topmost level, according to the accepted partition
  in Ref.~\cite{adamic_political_2005}}
\end{figure*}

In Fig.~\ref{fig:nested-pp} we show the results of the inference
procedure for a generated model with $B_0=2$ and $L=5$, $N=10^4$ nodes
and $\avg{k}=20$. The correct number of blocks is detected up to a given
value of $c > c^*$, where $c^*$ is the detectability threshold. The
hierarchy itself matches the nested PP model exactly only for higher
values of $c$, and becomes progressively simplified for lower
values. Note that for a large fraction of $c$ values the correct
lower-level partition is detected with a very high precision, but the
hierarchy that is inferred can be simpler than the planted one. In these
cases, however, both the inferred hierarchy, as well as the planted
model are fully equivalent, i.e. they generate the same networks [this
is true for the (a), (b) and (c) regions in Fig.~\ref{fig:nested-pp},
which have $B_0=16$]. In other words, the shallower hierarchies that are
inferred correspond to identical representations of the same $e_{rs}$
matrix at the lowest level, which require less information to be
described, in comparison to the sequence of Kronecker products used in
the model specification, and, hence, cannot really be seen as a failure
of the inference method, since it simply manages to compress the
original model. Before the value of $c$ reaches the detectability
threshold, the inference method settles on a fully random $L=1, B=1$
structure, corresponding once again to a parameter region where the
block detection is only possible with limited precision and if one knows
the correct model size. As predicated by the MDL criterion, the inferred
models tend to be as simple as possible, with the hierarchies becoming
shallower as one approaches a random graph. The approach is, therefore,
conservative, which brings confidence to the blocks and hierarchies
that are actually found, since despite the increased resolution
capabilities it does not tend to find spurious hierarchies.

In Appendix~\ref{app:benchmark} we also include a comparison of the
method with other algorithms for community detection that are not based
on statistical inference.

\section{Empirical Networks}

\begin{figure}
  \includegraphics[width=\columnwidth]{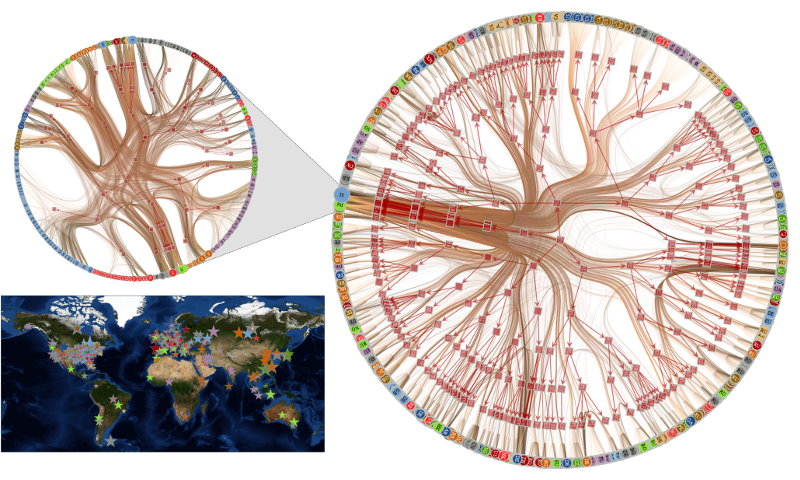} \caption{\label{fig:caida}
  Large-scale structure of the Internet at the autonomous systems level,
  as obtained by the nested stochastic block model, displaying a
  prominent core-periphery architecture. The magnification shows the
  nodes that belong to the ``core'' top-level branch, containing AS
  nodes spread all over the globe, as shown in the map inset. See the
  Supplemental Material for a higher-resolution version of this figure.}
\end{figure}

\begin{figure}
  \includegraphics[width=.8\columnwidth]{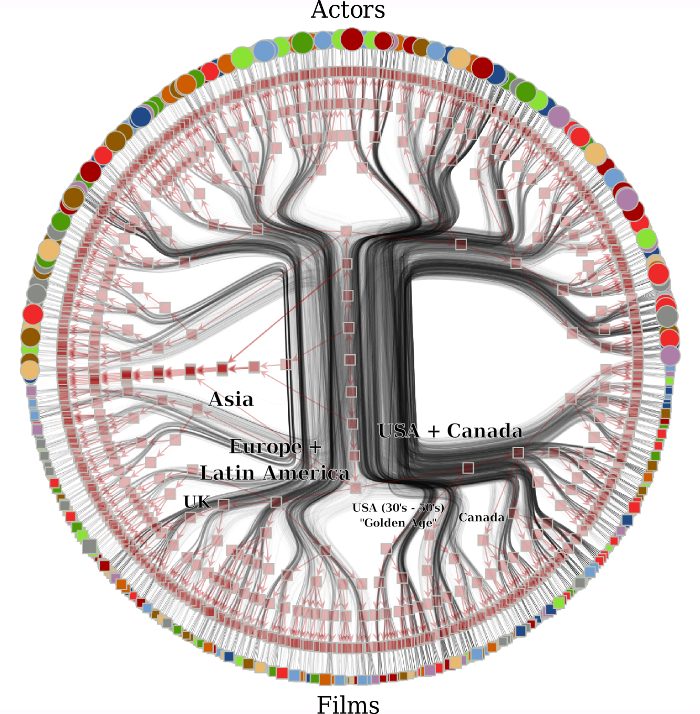} \caption{\label{fig:imdb}
  Large-scale structure of the IMDB film-actor network. Each node in
  this graph represents a lowest-level block in the hierarchy, instead
  of individual nodes in the graph. The size of the nodes indicates the
  number of nodes in each group. The hierarchy branch at the top are the
  actors, and at the bottom are the films. The labels classify each
  branch according to the most prominent geographical and temporal
  characteristics found in the database.  See the Supplemental Material
  for a higher-resolution version of this figure.}
\end{figure}

\begin{figure}[t]
  \begin{overpic}[width=.49\columnwidth]{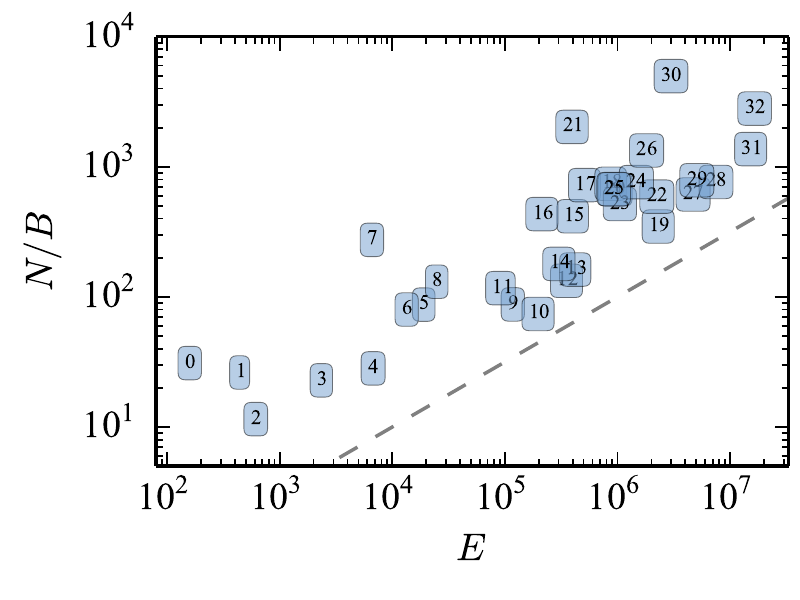}
    \put(24,61){\smaller (a)}
  \end{overpic}
  \begin{overpic}[width=.49\columnwidth]{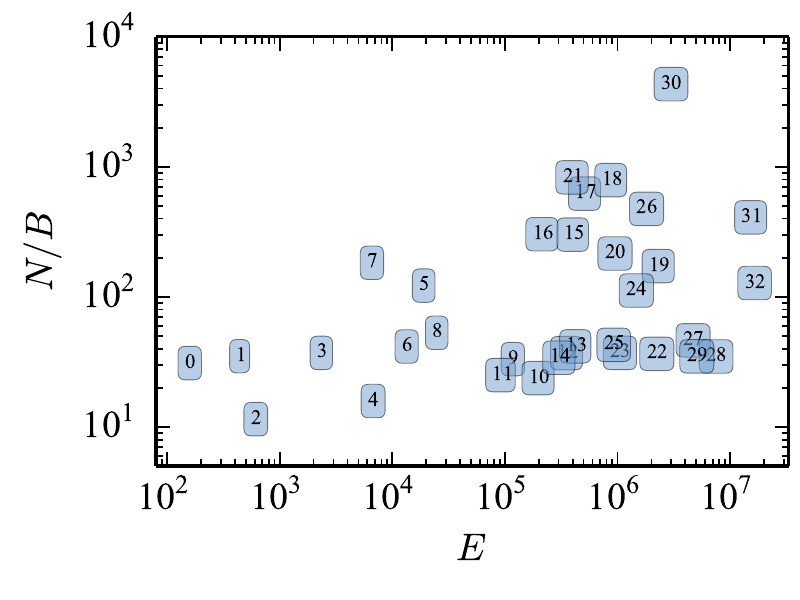}
    \put(24,61){\smaller (b)}
  \end{overpic}\\
  \begin{overpic}[width=.49\columnwidth]{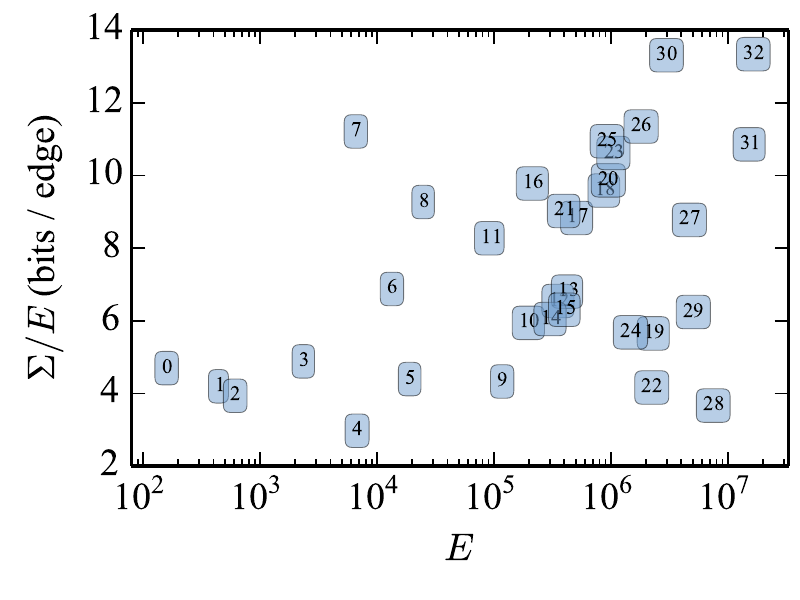}
    \put(20,63){\smaller (c)}
  \end{overpic}
  \begin{overpic}[width=.49\columnwidth]{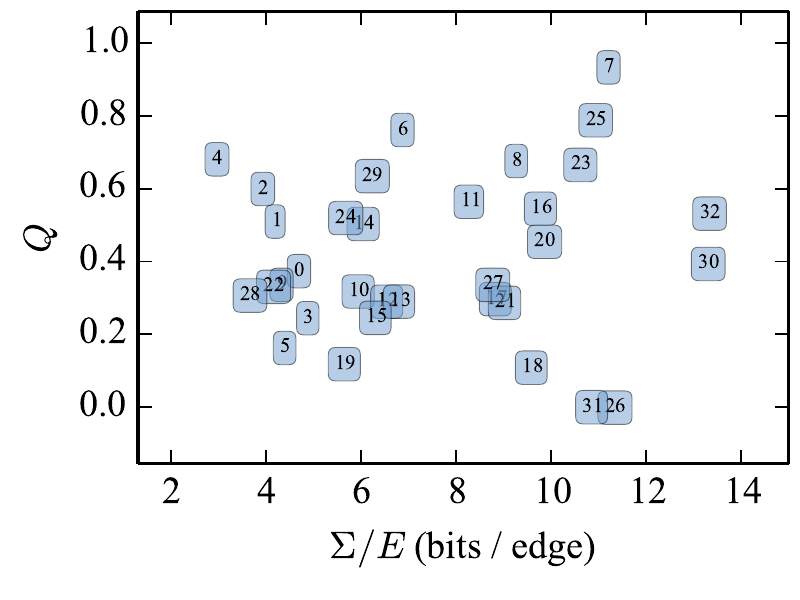}
    \put(22,65){\smaller (d)}
  \end{overpic}

  \resizebox{\columnwidth}{!}
            {\smaller[2]
              \setlength{\tabcolsep}{2pt}
              \begin{tabular}{l lll @{\hspace{8pt}}  l lll @{\hspace{8pt}}  l lll}\toprule
                No. & $N$ & $E$ &  Dir.  & No. & $N$ & $E$ & Dir.  & No. & $N$ & $E$ & Dir. \\ \midrule
0 & $62$ & $159$ & No  & 11 & $21,363$ & $91,286$ & No  & 22 & $255,265$ & $2,234,572$ & Yes \\ 
\rowcolor{gray!25}[2pt] 1 & $105$ & $441$ & No  & 12 & $27,400$ & $352,504$ & Yes  & 23 & $317,080$ & $1,049,866$ & No \\ 
2 & $115$ & $613$ & No  & 13 & $34,401$ & $421,441$ & Yes  & 24 & $325,729$ & $1,469,679$ & Yes \\ 
\rowcolor{gray!25}[2pt] 3 & $297$ & $2,345$ & Yes  & 14 & $39,796$ & $301,498$ & Yes  & 25 & $334,863$ & $925,872$ & No \\ 
4 & $903$ & $6,760$ & No  & 15 & $52,104$ & $399,625$ & Yes  & 26 & $372,547$ & $1,812,312$ & No \\ 
\rowcolor{gray!25}[2pt] 5 & $1,222$ & $19,021$ & Yes  & 16 & $56,739$ & $212,945$ & No  & 27 & $449,087$ & $4,690,321$ & Yes \\ 
6 & $4,158$ & $13,422$ & No  & 17 & $75,877$ & $508,836$ & Yes  & 28 & $654,782$ & $7,499,425$ & Yes \\ 
\rowcolor{gray!25}[2pt] 7 & $4,941$ & $6,594$ & No  & 18 & $82,168$ & $870,161$ & Yes  & 29 & $855,802$ & $5,066,842$ & Yes \\ 
8 & $8,638$ & $24,806$ & No  & 19 & $105,628$ & $2,299,623$ & No  & 30 & $1,134,890$ & $2,987,624$ & No \\ 
\rowcolor{gray!25}[2pt] 9 & $11,204$ & $117,619$ & No  & 20 & $196,591$ & $950,327$ & No  & 31 & $1,637,868$ & $15,205,016$ & No \\ 
10 & $17,903$ & $196,972$ & No  & 21 & $224,832$ & $394,400$ & Yes  & 32 & $3,764,117$ & $16,511,740$ & Yes \\ 
\bottomrule
              \end{tabular}
            }

            \vspace{2pt}
  \resizebox{\columnwidth}{!}
            {\smaller[2]
              \setlength{\tabcolsep}{2pt}
              \begin{tabular}{l l @{\hspace{8pt}}  l l @{\hspace{8pt}}  l l}
                No. & Network &  No. & Network & No. & Network \\ \midrule
0 & Dolphins~\cite{lusseau_bottlenose_2003}  & 11 & arXiv Co-Authors (cond-mat)~\cite{leskovec_graph_2007}  & 22 & Web graph of \url{stanford.edu}.~\cite{leskovec_community_2008} \\ 
\rowcolor{gray!25}[2pt] 1 & Political Books\footnote{V. Krebs, retrieved from~\url{http://www-personal.umich.edu/~mejn/netdata/}}  & 12 & arXiv Citations (hep-th)~\cite{leskovec_graph_2007, gehrke_overview_2003}  & 23 & DBLP collaboration~\cite{yang_defining_2012} \\ 
2 & American Football~\cite{girvan_community_2002, evans_american_2012}  & 13 & arXiv Citations (hep-ph)~\cite{leskovec_graph_2007, gehrke_overview_2003}  & 24 & WWW~\cite{albert_internet:_1999} \\ 
\rowcolor{gray!25}[2pt] 3 & C. Elegans Neurons~\cite{watts_collective_1998}  & 14 & PGP~\cite{richters_trust_2011}  & 25 & Amazon product network~\cite{yang_defining_2012} \\ 
4 & Disease Genes~\cite{goh_human_2007}  & 15 & Internet AS (Caida)\footnote{Retrieved from \url{http://www.caida.org}.}  & 26 & IMDB film-actor\footnote{Retrieved from \url{http://www.imdb.com/interfaces}.}~\cite{peixoto_parsimonious_2013} (bipartite) \\ 
\rowcolor{gray!25}[2pt] 5 & Political Blogs~\cite{adamic_political_2005}  & 16 & Brightkite social network~\cite{cho_friendship_2011}  & 27 & APS citations\footnote{Retrieved from \url{http://publish.aps.org/dataset}.} \\ 
6 & arXiv Co-Authors (gr-qc)~\cite{leskovec_graph_2007}  & 17 & Epinions.com trust network~\cite{richardson_trust_2003}  & 28 & Berkeley/Stanford web graph~\cite{leskovec_community_2008} \\ 
\rowcolor{gray!25}[2pt] 7 & Power Grid~\cite{watts_collective_1998}  & 18 & Slashdot~\cite{leskovec_signed_2010}  & 29 & Google web graph~\cite{leskovec_community_2008} \\ 
8 & arXiv Co-Authors (hep-th)~\cite{leskovec_graph_2007}  & 19 & Flickr~\cite{mcauley_image_2012}  & 30 & Youtube social network~\cite{yang_defining_2012} \\ 
\rowcolor{gray!25}[2pt] 9 & arXiv Co-Authors (hep-ph)~\cite{leskovec_graph_2007}  & 20 & Gowalla social network~\cite{cho_friendship_2011}  & 31 & Yahoo groups\footnote{Retrieved from \url{http://webscope.sandbox.yahoo.com}.} (bipartite) \\ 
10 & arXiv Co-Authors (astro-ph)~\cite{leskovec_graph_2007}  & 21 & EU email~\cite{leskovec_graph_2007}  & 32 & US patent citations~\cite{leskovec_graphs_2005} \\ 
\bottomrule
              \end{tabular}
            }
  \caption{\label{fig:meta} (a) The average block size $N/B$ obtained
  using the nonhierarchical model, as a function of $E$, for the
  empirical networks listed in the bottom table. The dashed line shows a
  $\sqrt{E}$ slope. (b) The same as (a) but with the nested model. (c)
  The description length $\Sigma/E$ for the nested model as a function
  of $E$. (d) The value of modularity $Q$ as function of $\Sigma/E$, for
  the nested model.}
\end{figure}

Here, we present a detailed analysis of some selected empirical networks,
as well as a meta-analysis of several networks, spanning different
domains and size scales. In all cases, we use the degree-corrected
stochastic block model at the lowest hierarchical level, instead of the
traditional model, since it almost always provides better results.

\noindent{\bfseries Political blogs of the 2004 US election.}  This is
a network compiled by Adamic et al~\cite{adamic_political_2005} of
political blogs during the 2004 presidential election in the USA. The
nodes are $N=1,222$ individual blogs, and $E=19,027$ directed edges
exist between pairs of blogs, if one blog cites the other. This network
is often used as an empirical example of community structure, since it
displays a division along political lines, with two clearly distinct
groups representing those aligned with the Republican and the Democratic
parties. Indeed, if one applies the nested block model to this network,
the topmost division in the hierarchy corresponds exactly to this
bimodal partition, which closely matches the accepted division (see
Fig.~\ref{fig:polblogs}). This partition is also obtained with the
nonhierarchical stochastic block model if one imposes
$B=2$~\cite{karrer_stochastic_2011}. However, the nested version reveals
a much more complete picture of the network, where these two partitions
possess a detailed internal structure, culminating in $B_0=15$ subgroups
with quite heterogeneous connection patterns. For instance, one can see
that each of the two higher-level groups possesses one or more subgroups
composed mainly of peripheral nodes, i.e., blogs that cite other blogs,
but are not themselves cited as often. Conversely, both factions possess
subgroups which tend to be cited by most other groups, and others which
are cited predominantly by specific groups. It is also interesting to
note that a large fraction of the connections between the two
top-level groups are concentrated between only two specific subgroups,
which, therefore, act as bridges between the larger groups.

This example shows that the model is capable of revealing the structure
of the network at multiple scales, which reveal simultaneously the
existence of the bimodal large-scale division, as well the lower-level
subdivisions.

\noindent{\bfseries The Autonomous Systems (AS) topology of the
Internet}. Autonomous Systems (AS) are intermediary building blocks of
the Internet topology. They represent organizational units that are used
to control the routing of packets in the network. A single AS often
corresponds to a network of its own, which is usually owned by a private
company, or a government body. The network analyzed here corresponds to
the traffic of information between the AS nodes, as measured by the
CAIDA project\footnote{The IPv4 Routed /24 AS Links Dataset,
\url{http://www.caida.org/data/active/ipv4_routed_topology_aslinks_dataset.xml}}. Each
node in the network is an AS, and a directed link exists between two
nodes if direct traffic has been observed between the two AS. As of
September 2013 the network is composed of $N=52,104$ AS nodes and
$E=399,625$ direct connections between them. The application of the
nested block model to this network yields the hierarchy seen in
Fig.~\ref{fig:caida}, with $B=191$ blocks at the lowest level. The most
prominent feature observed is a strong core-periphery structure, where
most connections go through a relatively small group of nodes, which act
as hubs in the network. The groups both in the core and in the periphery
seem strongly correlated to geographical location. However, the nodes of
the core groups are not confined to a single geographical location, and
are instead spread all over the globe (see inset of
Fig.~\ref{fig:caida}, and the Supplemental Material).

\noindent{\bfseries The Film-Actor Network.} This network is compiled by
extracting information available in the Internet Movie Database (IMDB),
which contains each cast member and film as distinct nodes, and an
undirected edge exists between a film and each of its cast members. If
nodes with a single connection are recursively removed, a network of
$N=372,447$ and $E=1,812,312$ remains (as of late 2012).  As can be seen
in Fig.~\ref{fig:imdb}, the nested block model fully captures the
bipartite nature of the network, and separates movies and actors at the
topmost hierarchical level, and proceeds to separate them in
geographical, temporal and topical (genre) lines. The observed partition
is similar to the one obtained via the nonhierarchical
model~\cite{peixoto_parsimonious_2013}, but one finds $B=971$ blocks,
instead of $B=332$ with the flat version.

\noindent{\bfseries Meta-analysis of several empirical networks.} We
perform an analysis of several empirical networks shown in
Fig.~\ref{fig:meta}, which belong to a wide variety of domains, and are
distributed across many size scales. We used the nonhierarchical
stochastic block model as well as the nested variant. In
Fig.~\ref{fig:meta}(a) and (b), we shown the average block sizes $N/B$
for all networks using both models. For the nonhierarchical version, a
clear $N/B \sim \sqrt{E}$ trend is observed, which corresponds to the
resolution limit present with this method, and other approaches as
well. In Fig.~\ref{fig:meta}(b) are shown the results for the nested
model, where such a trend can no longer be observed, and the smallest
average block sizes no longer seem to depend on the size of the network,
which serves as an empirical demonstration of the lack of resolution
limit shown previously. The values of the description lengths themselves
are also distributed in a seemingly nonorganized manner [see
Fig.~\ref{fig:meta}(c)], i.e. no general tendency for larger networks
can be observed, other than an increased range of possible values for
larger $E$ values. Any difference observed seems to be due to the actual
topological organization, rather than intrinsic constraints imposed by
the method.  We also compute the modularity of the inferred block
structures, $Q=\sum_{r}e_{rr}/2E - e_r^2/(2E)^2$, which measures how
assortative the topology is. Higher values of $Q$ close to $1$ indicate
the existence of densely connected communities. The value of $Q$ is the
most common quantity used to detect blocks in networks, and it presumes
that such assortative connections are present. In contrast, by fitting a
general stochastic block model, no specific pattern is assumed, and the
partition found corresponds to the least random model that matches the
data. In Fig.~\ref{fig:meta} we show the values of $Q$ obtained for
the analyzed networks. Indeed, some networks are modular, with high
values of $Q$. However, one does not observe any strong correlation of
the description length and the modularity values. Hence, the most
structured networks do not necessarily possess much larger $Q$ values,
which indicate that the building blocks of their topological
organization are not predominantly assortative communities (this is
clear in some of the examples considered previously, such as the
Internet AS topology and the IMDB network). However, for many of these
networks, it is probably possible to find partitions that lead to much
higher $Q$ values. These partitions would, on the other hand, correspond
to block model ensembles with a larger entropy than those inferred via
maximum likelihood. Therefore, the maximization of $Q$ in these cases
would invariably discard topological information present in the network,
and provide a much simplified and possibly misleading picture of the
large-scale structure of the network. Hence, it seems more appropriate to
confine modularity maximization only to cases where the assortative
structure is known to be the dominating pattern. However, even in these
cases, methods based on statistical inference possess clear advantages,
such as the lack of resolution limit, model selection guarantees, and
the overall more principled nature of the approach.

\section{Discussion}

In this paper, we present a principled method to detect hierarchical
structures in networks via a nested stochastic block model. This method
fully generalizes previous approaches for the detection of hierarchical
community structures~\cite{clauset_structural_2007,
clauset_hierarchical_2008, rosvall_multilevel_2011,
sales-pardo_extracting_2007,
ronhovde_multiresolution_2009,kovacs_community_2010,park_dynamic_2010},
since it makes no assumptions either on the actual types of large-scale
structures possible (assortative, dissortative, or any arbitrary
mixture), or on the hierarchical form, which is not confined to binary
trees or dendograms. We show that a major advantage of this approach is
that it breaks the so-called resolution limit of approaches, such as
modularity optimization and nonhierarchical model inference, where
modules smaller than a characteristic size scaling with $\sqrt{N}$
cannot be resolved. With the nested model presented, this characteristic
scale is replaced by a much smaller logarithmic dependence, making it,
in practice, non-existent for many applications. This increased
resolution comes as a result of robust model selection principles, and
is integrated with the desirable capacity of differentiating between
noise and actual structure, and, therefore, it is not susceptible to the
detection of spurious communities.  We show that the model is capable of
inferring the large-scale features of empirical networks in significant
detail, even for very large networks.

This type of approach should, in principle, also be applicable to other
model classes, such as those based on
overlapping~\cite{palla_uncovering_2005, airoldi_mixed_2008,
lancichinetti_detecting_2009, gopalan_efficient_2013}, or link
communities~\cite{ahn_link_2010, ball_efficient_2011}. We also predict
that it should serve as a more refined method of detecting missing
information in networks~\cite{clauset_hierarchical_2008,
guimera_missing_2009}, as well as for the prediction of the network
evolution~\cite{liben-nowell_link-prediction_2007}, determining the more
salient topological features~\cite{grady_robust_2012,
bianconi_assessing_2009}, or large-scale functional summaries of the
network topology~\cite{guimera_functional_2005}.

\begin{acknowledgments}
The author would like to thank Sebastian Krause for a careful reading of
the manuscript, as well as Cris Moore and Lenka Zdeborová for useful
comments. This work was funded by the University of Bremen, under the
funding line ZF04.
\end{acknowledgments}

\appendix
\section{Bayesian model selection (BMS)}\label{app:bms}

In the following we compare Bayesian model selection (BMS) via
integrated likelihood with the MDL approach considered in the main text,
and we show that they lead to the same criterion if the model
constraints are equivalent.

For the purpose of performing BMS, we evoke the most usual definition of
the stochastic block model ensemble, where one defines as parameters the
probabilities $p_{rs}$ that an edge exists between two nodes belonging
to blocks $r$ and $s$. The posterior likelihood of observing a given
graph with a block partition $\{b_i\}$ and model parameters $\{p_{rs}\}$
is
\begin{equation}\label{eq:l}
  \mathcal{P}(G|\{b_i\}, \{p_{rs}\}, B) = \prod_{rs}p_{rs}^{\frac{e_{rs}}{2}}(1-p_{rs})^{\frac{n_rn_r-e_{rs}}{2}}.
\end{equation}
The inference procedure consists in, as before, maximizing this quantity
with respect to the parameters $\{p_{rs}\}$ and the block partition
$\{b_i\}$. It is easy to see that if one maximizes Eq.~\ref{eq:l} with
respect to $\{p_{rs}\}$, one recovers
$\max_{\{p_{rs}\}}\ln\mathcal{P}(G|\{b_i\}, \{p_{rs}\}, B) =
-\mathcal{S}_t$, given in Eq.~\ref{eq:st}, so indeed these models are
equivalent. However this does not provide a means for model selection,
since models with a larger number of blocks $B$ will invariably posses a
larger likelihood. Instead, the Bayesian model selection approach is to
consider the joint probability $\mathcal{P}(G, \{b_i\}, \{p_{rs}\},
\{p_r\} | B)$ of observing not only the graph, but also the partition
$\{b_i\}$, the model parameters $\{p_{rs}\}$ as well as the parameters
$\{p_r\}$ that control the probability of each partition $\{b_i\}$
being observed, which is given by
\begin{equation}\label{eq:bprob}
  \mathcal{P}(\{b_i\}| \{p_r\}, B) = \prod_rp_r^{n_r}.
\end{equation}
This invariably leads to the inclusion of prior probabilities of
observing the model parameters, $\mathcal{P}(\{p_{rs}\}|B)$ and
$\mathcal{P}(\{p_r\}|B)$. Now, instead of finding the model parameters
that maximize this quantity, we compute the \emph{integrated
likelihood}~\cite{biernacki_assessing_2000,daudin_mixture_2008,
come_model_2013},
\begin{align}
  \mathcal{P}(G,\{b_i\}| B) &= \!\int\!\mathrm{d}p_{rs}\mathrm{d}p_r\,\mathcal{P}(G, \{b_i\}, \{p_{rs}\}, \{p_r\} | B) \label{eq:il}\\
  \begin{split}
    =\int\!\mathrm{d}p_{rs}\,\mathcal{P}(G| \{b_i\}, \{p_{rs}\}, B)\mathcal{P}(\{p_{rs}\}|B)\times\\
    \int\!\mathrm{d}p_r\mathcal{P}(\{b_i\}|\{p_r\})\mathcal{P}(\{p_r\} | B)  \label{eq:il_int}
  \end{split}\\
    &= \mathcal{P}(G|\{b_i\},B)\times\mathcal{P}(\{b_i\}|B).\label{eq:il_summary}
\end{align}
By maximizing $\mathcal{P}(G,\{b_i\}| B)$, instead of Eq.~\ref{eq:l},
one should avoid overfitting the data, since the larger models with many
parameters are dominated by a majority of choices that fit the data very
badly, and, hence, have a smaller contribution in the integral of
Eq.~\ref{eq:il}. Therefore, the maximization of the integrated
likelihood also corresponds to an application of Occam's razor, and one
should expect it to deliver results compatible with
MDL~\cite{grunwald_minimum_2007}. However, in practice things are more
nuanced, since the value of Eq.~\ref{eq:il} is heavily dependent on the
choice of priors $\mathcal{P}(\{p_{rs}\}|B)$ and
$\mathcal{P}(\{p_r\}|B)$. For the block partitions themselves, this
choice is more straightforward. Since one wants to be agnostic with
respect to what block sizes are possible, one should choose a flat prior
$\mathcal{P}(\{p_r\}|B) = \operatorname{Dirichlet}(\{p_r\}|\{\alpha_r\})$, with
$\alpha_r=1$, so that all counts are equally likely. The integral of
Eq.~\ref{eq:il_int} is then computed as
\begin{equation}
  \ln\mathcal{P}(\{b_i\}|B) = - \ln{\textstyle \multiset{B}{N}} - \ln N! + \sum_r \ln n_r!,
\end{equation}
which is identical to the partition description length of
Eq.~\ref{eq:dli}, i.e. $\ln\mathcal{P}(\{b_i\}|B) =
-\mathcal{L}^0_t$.

For the block probabilities, on the other hand, the situation is more
subtle. A common choice is the flat prior $\mathcal{P}(\{p_{rs}\}|B) =
1$~\cite{guimera_missing_2009, daudin_mixture_2008, moore_active_2011,
latouche_variational_2012, come_model_2013}. This choice is agnostic with
respect to what block structures are expected, and it is also practical,
since the integral can be evaluated exactly~\cite{guimera_missing_2009,come_model_2013},
\begin{gather}
  \begin{split}
    \ln\mathcal{P}(G|\{b_i\},B) = -\sum_{r>s}\ln{{n_{r}n_{s}} \choose e_{rs}}+\ln\left(n_rn_s + 1\right) \\
    -\sum_r\ln{n_{r}^2 \choose e_{rr}/2}+\ln\left(n_r^2/2+ 1\right)\\
  \end{split}\\
  \simeq -\frac{1}{2}\sum_{rs}n_rn_sH_{\text{b}}\left(\frac{e_{rs}}{n_rn_s}\right) -(B+1)\sum_r\ln n_r\label{eq:bms_dense},
\end{gather}
where the approximation in Eq.~\ref{eq:bms_dense} was made assuming
$n_r\gg 1$, and $H_{\text{b}}(x)$ is the binary entropy function. However, there is
one important issue with this approach. Namely, there is a strong
discrepancy between the models generated by the flat prior
$\mathcal{P}(\{p_{rs}\}|B) = 1$ and most observed empirical
networks. Specifically, typical parameters with $p_{rs} = 1/2$ sampled
by this prior will result in \emph{dense} networks with average degree
$\avg{k}=\sum_{rs}p_{rs}n_rn_s/N=N/2$.  However, most large empirical
networks tend to be \emph{sparse}, with an average degree which is many
orders of magnitude smaller than $N$. Hence, as $N$ becomes large, most
observed networks will lie in a vanishingly small portion of the
parameter space produced by this prior. A better choice would constrain
the average degree to something closer to what is observed in the data,
but at the same time being otherwise noninformative regarding the block
structure. A choice such as $\mathcal{P}(\{p_{rs}\}|B) \propto
\delta(\sum_{rs}p_{rs}n_rn_s -2E)$, where $E$ is the number of edges in
the observed network seems appropriate, but the integral in
Eq.~\ref{eq:il_int} becomes difficult to solve. Instead, an easier
approach is to modify the model sightly, so that the average degree is
implicitly constrained. Here, we consider the model variant where the
number of edges $E$ is a fixed parameter, and each sampled edge may land
between any two nodes belonging to blocks $r$ and $s$ with probability
$q_{rs}$, and we have, therefore, $\sum_{r\ge s}q_{rs} = 1$. The full
posterior likelihood of this model is
\begin{equation}\label{eq:l_sparse}
  \mathcal{P}(G|\{b_i\}, \{q_{rs}\}, E, B) = \frac{E!}{\Omega(\{e_{rs}\},\{n_{r}\})} \frac{\prod_{r \ge s} q_{rs}^{m_{rs}}}{\prod_{r\ge s}m_{rs}!},
\end{equation}
where $\Omega(\{e_{rs}\}, \{n_{r}\})$ is, as before, the number of
different graphs with the same block partition and edge counts, and
$m_{rs} = e_{rs}$ if $r\neq s$ or $e_{rr} /2$ otherwise. By maximizing
Eq.~\ref{eq:l_sparse} with respect to $\{q_{rs}\}$, one obtains
$\max_{\{q_{rs}\}}\ln\mathcal{P}(G|\{b_i\}, \{q_{rs}\}, E, B) \simeq
-\ln\Omega(\{e_{rs}\},\{n_{r}\})=-\mathcal{S}_t$, as long as $m_{rs}\gg
1$ or $m_{rs}=0$, so it also is equivalent to the previous models in
this limit. With this reparametrization, the average degree remains
fixed independently of the choice of prior. Therefore, we may finally use
a flat prior $\mathcal{P}(\{q_{rs}\}|B) =
\operatorname{Dirichlet}(\{q_{rs}\}|\{\alpha_{rs}=1\})$, without the risk of the
graphs becoming inadvertently dense, and again the integrated likelihood
can be computed exactly,
\begin{align}
  \mathcal{P}(G|\{b_i\},B) &= \int\!\mathrm{d}q_{rs}\,\mathcal{P}(G| \{b_i\}, \{q_{rs}\}, E, B)\mathcal{P}(\{q_{rs}\}|B)\\
  &=\left[\Omega(\{e_{rs}\},\{n_{r}\})\times{\textstyle \multiset{\multiset{B}{2}}{E}}\right]^{-1} \label{eq:bms_sparse}.
\end{align}
By inserting Eq.~\ref{eq:bms_sparse} into Eq.~\ref{eq:il_summary}, and
comparing with equation Eq.~\ref{eq:dl_flat}, we see
that $\ln\mathcal{P}(G,\{b_i\}| B) = -\Sigma_{L=1}$, and we conclude
reassuringly that the MDL approach is fully equivalent to BMS when all
model constraints are compatible. In fact, even in the dense case,
although not quite the same, the (dense) BMS and MDL penalties are very
similar. If one assumes $N \gg B^2$, $E\propto N^2$, and equal block
sizes $n_r=N/B$, both penalties become $\sim B(B+1) \ln N + N\ln
B$. Therefore, it seems that whatever differences arising from the two
approaches stem simply from nuances in the choice of prior
probabilities. This comparison also allows us to interpret the nested
block model as a hierarchical Bayesian approach, where the priors
$\mathcal{P}(\{q_{rs}\}|B)$ are replaced by a nested sequence of priors
and hyperpriors, so that their integrated likelihood matches the
description length defined previously.

\section{Comparison with other community detection methods}\label{app:benchmark}

In this section we compare results obtained for synthetic networks with
popular community detection methods that are not based on statistical
inference. Here, we focus not only on the capacity of the method of
finding a partition correlated with the planted one, but also on the
number of blocks detected. We concentrate on two methods which have been
reported to provide good results in synthetic
benchmarks~\cite{lancichinetti_community_2009}, namely the Louvain
method~\cite{blondel_fast_2008}, based on modularity optimization, and
the Infomod method~\cite{rosvall_maps_2008, rosvall_map_2009,
rosvall_multilevel_2011}, based on compression of random walks. We make
use of the LFR benchmark~\cite{lancichinetti_benchmark_2008}, which
corresponds to a specific parametrization of the degree-corrected
stochastic block model~\cite{karrer_stochastic_2011}, where both the
degree distribution and the block size distribution follow truncated
power laws. Here, we employ a parametrization similar to
Ref.~\cite{lancichinetti_community_2009}, with a degree distribution
following a power law with exponent $-2$ and a minimum degree
$k_{\text{min}}=5$, and a community size distribution also following a
power-law, but with exponent $-1$, and minimum block size of $50$. We
also impose the following additional restrictions: The total number of
blocks is always fixed at $B=100$, and for every node $i$ belonging to
block $r$, its degree $k_i$ cannot exceed $\sqrt{n_r}$, to avoid
intrinsic degree-degree correlations~\cite{peixoto_entropy_2012}. With
this parameter choice, the networks generated with $N=2\times 10^4$
possess an average degree $\avg{k}\simeq 7.8$. The actual block
structure is parametrized as $e_{rs} = (1-c) e_re_s / 2E + \delta_{rs}c
e_r$, where $c$ controls the assortativity: For $c=1$ all edges connect
nodes of the same block, and for $c=0$ we have a fully random
configuration model\footnote{Note that this is slightly different than
in Ref.~\cite{lancichinetti_benchmark_2008}, which parameterized the
fraction of internal and external degrees via a local mixing parameter
$\mu$, which is the same for all communities. That choice corresponds to
a different parametrization of the degree-corrected block model than the
one used here. However, since the blocks have different sizes, and the
degrees are approximately the same in all blocks, in general there is no
choice of $\mu$ which would allow one to recover the fully-random
configuration model, since the intrinsic mixing would be different for
each block in this case. Because of this, we have opted for the
parametrization used here, however this should not alter the
interpretation of the benchmark and the comparison with
Ref.~\cite{lancichinetti_benchmark_2008} in a significant way.}.

Because the different methods result in quite different numbers of
detected blocks, the normalized mutual information (NMI) is not the most
appropriate measure of the overlap between partitions in this case. This
is due to the fact that, if the number of nodes is kept fixed, the NMI
values tend to be larger simply if the number of blocks is increased,
even if this larger partition is in no other way more strongly
correlated to the true one. Another measure that is less susceptible to
this problem is the variation of information
(VI)~\cite{meila_comparing_2003}, defined as
\begin{equation}
  \text{VI}(\{x_i\}, \{y_i\}) = H(\{x_i\}) + H(\{y_i\}) - 2I(\{x_i\},\{y_i\}),
\end{equation}
where $H(\{x_i\})$ is the entropy of the partition $\{x_i\}$ and
$I(\{x_i\},\{x_i\})$ is the (non-normalized) mutual information between
$\{x_i\}$ and $\{y_i\}$. A value of VI equal to zero means that the
partitions are identical, whereas any positive value indicates a reduced
overlap between them.

\begin{figure}[t]
  \includegraphics[width=0.95\columnwidth]{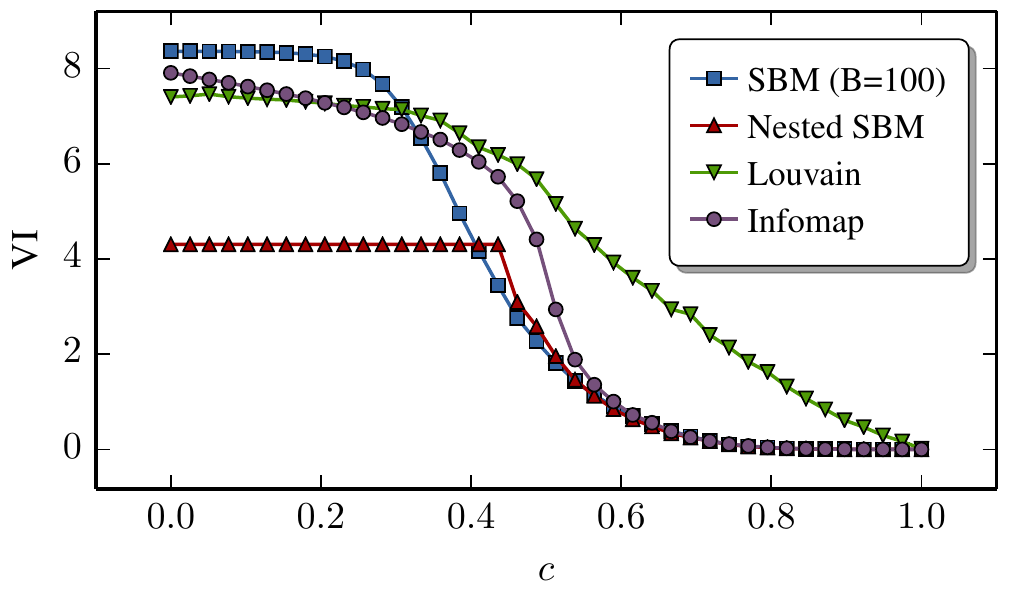}
  \includegraphics[width=0.95\columnwidth]{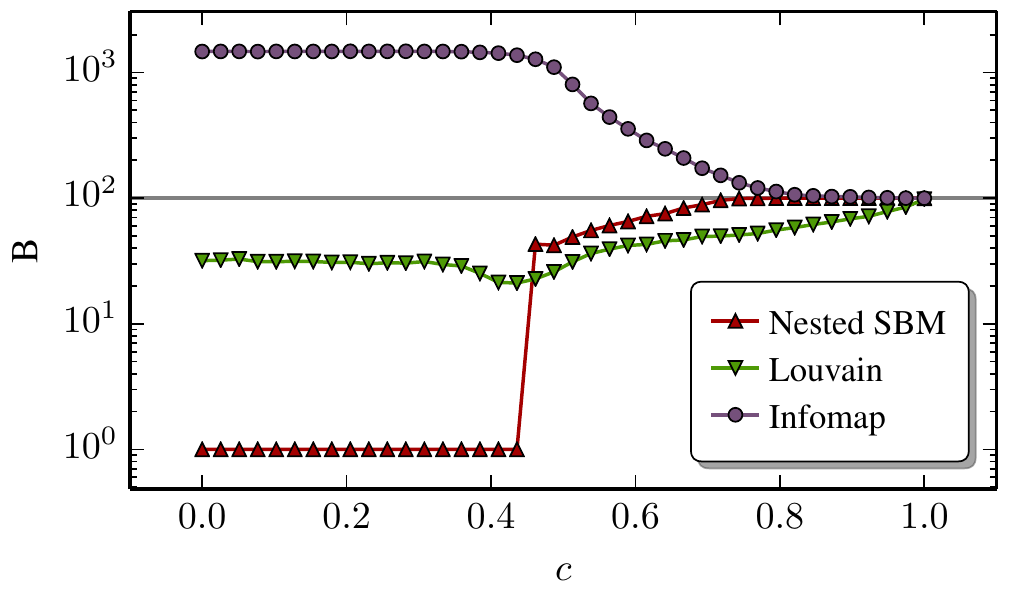}

  \caption{\label{fig:comparison} \emph{Top:} Variation of information
  (VI) between the planted and obtained partitions as a function of the
  assortativity parameter $c$, for networks with $N=2\times 10^4$,
  generated as described in the text. The legend indicates results
  obtained with different methods: Fitting the degree-corrected
  stochastic block model with a fixed number of blocks $B=100$ (SBM),
  performing model selection with the nested stochastic block model
  (Nested SBM), the Louvain modularity maximization
  method~\cite{blondel_fast_2008}, and the Infomod
  method~\cite{rosvall_maps_2008, rosvall_map_2009,
rosvall_multilevel_2011}. \emph{Bottom:} The obtained number of blocks
$B$ as a function of $c$, for the same methods as in the top panel. The
gray horizontal line marks the planted $B=100$ value. All results were
obtained by averaging over $20$ network realizations.}
\end{figure}

The VI values between the planted partitions and those obtained with
different methods for several network realizations of the above model
are shown in Fig.~\ref{fig:comparison}, together with the obtained
number of blocks. By observing the VI values for the inference method
with a fixed number of blocks $B=100$, we conclude that the strict
detectability transition (when the value of $B$ is known) lies somewhere
slightly above $c\approx 0.2$. However, the model-selection procedure
based on the nested stochastic block model presented in the main text
discards any structure below the $c\approx 0.4$ range, and decides on a
fully random $B=1$ structure. Above this value, the inferred value of
$B$ increases from $B=1$ until agreeing with the planted value for
sufficiently large $c$ values.  As can also be seen in
Fig.~\ref{fig:comparison}, the Louvain method exhibits the ``worst of
both worlds,'' i.e., it fails to find the correct partition for all
values except $c=1$, finding systematically smaller values of $B$, while
at the same time finding spurious partitions below the detectability
threshold, even when the network is completely random ($c=0$). The
Infomod method, on the other hand, seems to find partitions that are
largely compatible with the planted one, at least for the parameter
region above $c\approx 0.6$. However, for even larger values of $c$, this
method detects a number of blocks that is significantly larger than the
planted value, which increases steadily as $c$ decreases. Hence, this
method is also incapable of separating structure from noise, and finds
spurious partitions far below the detectability threshold. Thus, from
the three methods analyzed, the one described in the main text is the
only one that combines the following three desirable properties:
1. Optimal inference in the detectable range; 2. Guarantee against
overfitting and detection of spurious modules; 3. Fully nonparametric
implementation.

The suboptimal behavior of the modularity-based method is simply a
combination of the resolution limit~\cite{fortunato_resolution_2007} and
lack of built-in model selection based on statistical
evidence~\cite{guimera_modularity_2004}. It is not currently known if
the Infomod method suffers from problems similar to the resolution
limit, but clearly it lacks guarantees against detection of spurious
modules. Although it is also based on the principle of parsimony, it
tries to compress random walks taking place on the network, instead of
the network itself. Apparently, the method cannot distinguish between
the actual planted block structure and quenched topological fluctuations
--- both of which will affect random walks --- and gradually transitions
between the two properties in order to best describe the network
dynamics. (As has been shown in
Ref.~\cite{lancichinetti_community_2009}, this problem diminishes if the
average degree of the network is made sufficiently large, in which case
the method finally settles in a $B=1$ partition for fully random
graphs.)  On the other hand, the method in the main text is based on
maximizing the likelihood of the \emph{exact same} generative process
that was used to construct the network, which puts it in clear
advantage over the other two (and, in fact, many other methods, including
all those analyzed in Refs.~\cite{lancichinetti_benchmark_2008,
lancichinetti_community_2009}), in addition to including a robust and
formally motivated model-selection procedure.

\section{Directed and undirected networks}\label{app:summary}

As mentioned in the main text, the model described is easily generalized
for directed graphs. For the ensemble entropies, we have for the
undirected case~\cite{peixoto_entropy_2012},
\begin{equation}\label{eq:st_a}
  \mathcal{S}_t = \frac{1}{2} \sum_{rs}n_rn_sH_{\text{b}}\left(\frac{e_{rs}}{n_rn_s}\right),
\end{equation}
while for the directed case it reads,
\begin{equation}\label{eq:std_a}
  \mathcal{S}^d_t = \sum_{rs}n_rn_sH_{\text{b}}\left(\frac{e_{rs}}{n_rn_s}\right),
\end{equation}
where $H_{\text{b}}(x) = -x\ln x - (1-x)\ln(1-x)$ is the binary entropy
function. In both cases, $e_{rs}$ is the number of edges from block $r$
to $s$ (or the number of half-edges for the undirected case when $r=s$),
and $n_r$ is the number of nodes in block $r$. In the sparse limit,
$e_{rs}\ll n_rn_s$, these expressions may be written approximately as
\begin{align}
  \mathcal{S}_t &\cong E - \frac{1}{2} \sum_{rs}e_{rs}\ln\left(\frac{e_{rs}}{n_rn_s}\right),  \label{eq:st_s_a}\\
  \mathcal{S}^d_t &\cong E - \sum_{rs}e_{rs}\ln\left(\frac{e_{rs}}{n_rn_s}\right).   \label{eq:std_s_a}
\end{align}
For the degree-corrected variant with ``hard'' degree constraints, we
have
\begin{align}
  \mathcal{S}_c &\cong -E -\sum_kN_k\ln k! - \frac{1}{2} \sum_{rs}e_{rs}\ln\left(\frac{e_{rs}}{e_re_s}\right), \label{eq:sc_a}\\
  \begin{split}
  \mathcal{S}^d_c &\cong -E -\sum_{k^+}N_{k^+}\ln k^+!  -\sum_{k^-}N_{k^-}\ln k^-! \\
  & \qquad - \sum_{rs}e_{rs}\ln\left(\frac{e_{rs}}{e^+_re^-_s}\right),\label{eq:scd_a}
  \end{split}
\end{align}
where $e_r = \sum_se_{rs}$ is the number of half-edges incident on block
$r$, and $e^+_r = \sum_se_{rs}$ and $e^-_r = \sum_se_{sr}$ are the
number of out- and in-edges adjacent to block $r$, respectively. These
expressions are also only valid in the sparse limit, which in this case
amounts to the following conditions,
\begin{equation}
e_{rs}\frac{\avg{k^2}_r - \avg{k}_r}{\avg{k}^2_r}\frac{\avg{k^2}_s - \avg{k}_s}{\avg{k}^2_s} \ll n_rn_s,
\end{equation}
where $\avg{k^l}_r = \sum_{i\in r}k_i^l/n_r$ [for the directed case we
simply replace $\avg{k^l}_r \to \avg{(k^+)^l}_r$ and $\avg{k^l}_s \to
\avg{(k^-)^l}_s$ in the equation above]. Unfortunately, there is no
closed-form expression for the entropy outside the sparse limit, unlike
the traditional variant~\cite{peixoto_entropy_2012}.

For the upper-level multigraphs the entropies
are~\cite{peixoto_entropy_2012},
\begin{align}\label{eq:sm_a}
  \mathcal{S}_m  &= \sum_{r>s} \ln{\textstyle \multiset{n_rn_s}{e_{rs}}} + \sum_r \ln{\textstyle \multiset{\multiset{n_r}{2}}{e_{rr}/2}},\\
  \mathcal{S}^d_m  &= \sum_{rs} \ln{\textstyle \multiset{n_rn_s}{e_{rs}}},
\end{align}
where, as before, $\multiset{n}{m}={n+m-1\choose m}$ is the number of
$m$-combinations with repetitions from a set of size $n$.

For the degree-corrected model, the description length needs
to be augmented with the information necessary to describe the degree
sequence, analogously to Eq.~\ref{eq:l_c} for the undirected case,
\begin{equation}\label{eq:l_c_d}
  \mathcal{L}_c =  \mathcal{L}_t + \sum_rn_rH(\{p^r_{k^-,k^+}\}),
\end{equation}
where $\{p^r_{k^-,k^+}\}$ is the joint (in,out)-degree distribution of
nodes belonging to block $r$.

Note that other generalizations for the directed case are
possible~\cite{zhu_oriented_2014}, and it should be straightforward to
adapt the nested model for them as well.

\bibliography{bib}

\newpage
\onecolumngrid
\includepdf[pages=1,delta=0 2\textheight]{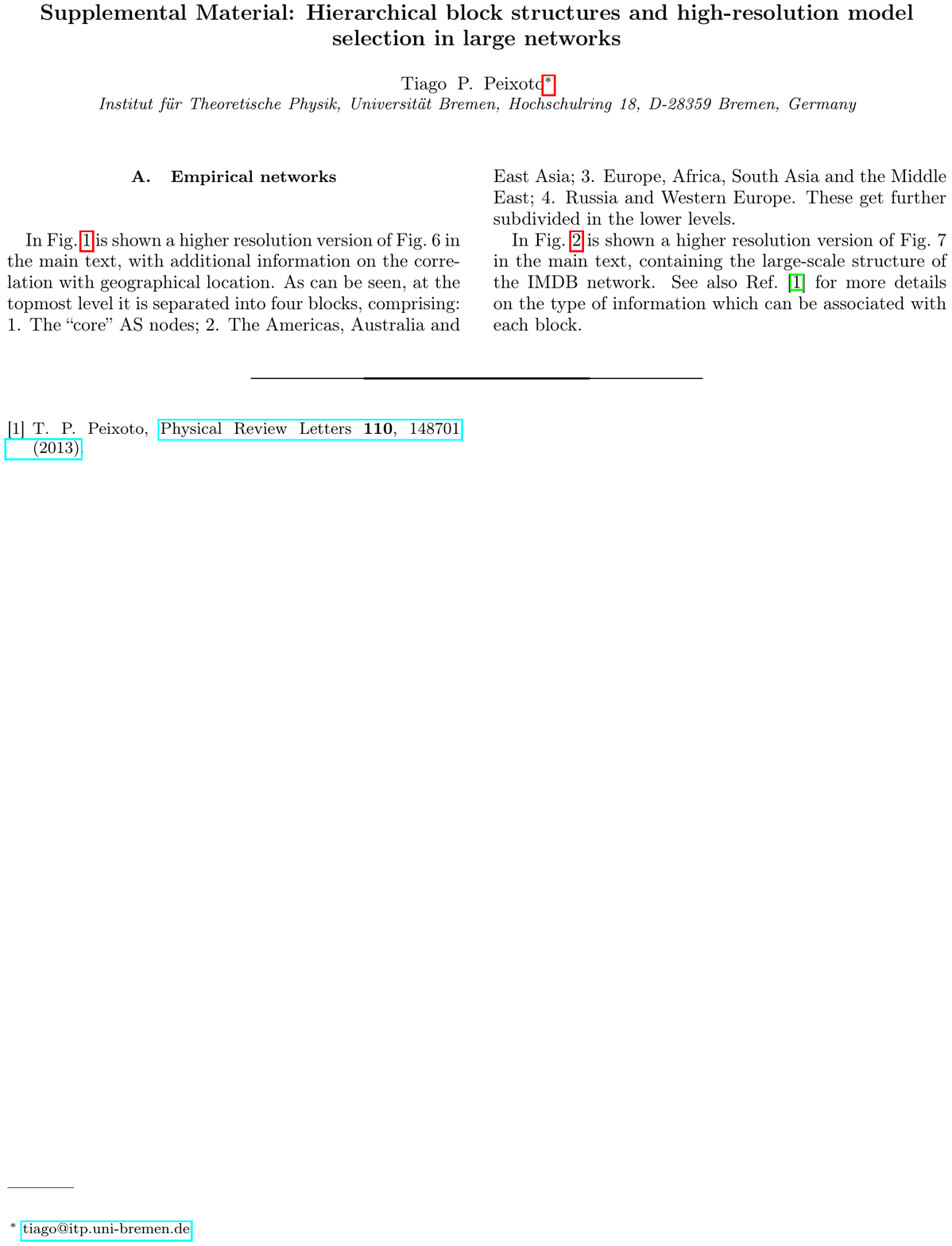}
\includepdf[pages=2,delta=0 2\textheight]{sup_inf.pdf}
\includepdf[pages=3,delta=0 2\textheight]{sup_inf.pdf}

\end{document}